\documentclass{article}

\usepackage{complex-systems}

\pdfoutput=1

\usepackage{graphics}
\usepackage{graphicx}
\usepackage[centertags]{amsmath}
\usepackage{amsfonts}
\usepackage{amssymb}
\usepackage{url}
\usepackage{subfigure}

\begin{document}

\title{Complete Characterization of Structure of Rule 54}

\author{
\authname{Genaro J. Mart{\'i}nez}
\\ [2pt]
\authadd{Escuela Superior de C\'omputo, Instituto Polit\'ecnico Nacional, M\'exico}\\
\authadd{Unconventional Computing Center, Computer Science Department,}\\ 
\authadd{University of the West of England, Bristol BS16 1QY, United Kingdom}\\
\authadd{\url{genaro.martinez@uwe.ac.uk}}%
\\ [2pt]
\and
\authname{Andrew Adamatzky}
\\[2pt] 
\authadd{Unconventional Computing Center, Computer Science Department,}\\ 
\authadd{University of the West of England, Bristol BS16 1QY, United Kingdom}\\ 
\authadd{\url{andrew.adamatzky@uwe.ac.uk}}%
\\[2pt]
\and
\authname{Harold V. McIntosh}
\\[2pt] 
\authadd{Departamento de Aplicaci\'on de Microcomputadoras,}\\ 
\authadd{Instituto de Ciencias, Universidad Aut\'onoma de Puebla, Puebla, M\'exico}\\ 
\authadd{\url{mcintosh@unam.mx}}%
}


\maketitle

\markboth{G. J. Mart{\'i}nez, A. Adamatzky, and H. V. McIntosh}{Complete Characterization of Structure of Rule 54}

\begin{abstract}
The dynamics of rule 54 one-dimensional two-state cellular automaton (CA) are a discrete analog of a space-time dynamics of excitations in nonlinear active medium with mutual inhibition. A cell switches its state 0 to state 1 if one of its two neighbors is in state 1 (propagation of a perturbation) and a cell remains in state 1 only if its two neighbors are in state 0. A lateral inhibition is because a 1-state neighbor causes a 1-state cell to switch to state 0. The rule produces a rich spectrum of space-time dynamics, including gliders and glider guns just from four primitive gliders. We construct a catalogue of gliders and describe them by tiles. We calculate a subset of regular expressions $\Psi_{R54}$ to encode gliders. The regular expressions are derived from de Bruijn diagrams, tile-based representation of gliders, and cycle diagrams sometimes. We construct an abstract machine that recognizes regular expressions of gliders in rule 54 and validate $\Psi_{R54}$. We also propose a way to code initial configurations of gliders to depict any type of collision between the gliders and explore self-organization of gliders, formation of larger tiles, and soliton-like interactions of gliders and computable devices.
\end{abstract}

\mbox{} \\

\noindent Paper published in {\em Complex Systems}, October, 2014. \\\url{http://www.complex-systems.com/abstracts/v23_i03_a04.html}.

\section{Preliminaries}

Cellular automata (CAs) are renowned for the simplicity of their rules and the complexity of their space-time evolution. Rule 54 is among the most famous rules which exhibit a nontrivial space-time dynamics. The rule belongs to complexity class IV in Wolfram's classification~\cite{kn:Wolf94,kn:Mar13}.

Rule 54 has always attracted considerable interest from computer scientists, mathematicians, and physicists, and thus, compared to other elementary cellular automaton (ECA) rules, is well investigated. Boccara et al. \cite{kn:BNR91} enumerated a number of gliders in rule 54 and characterized a glider gun. They applied statistical analysis to study the stability of gliders. Hanson and Crutchfield \cite{kn:HC97} introduced a concept of ``computational mechanics'', or designing of finite-state machines derived from language representations and motion equations of filtered gliders. Another exploration of rule 54 with automatic filters was presented by Wuensche in \cite{kn:Wue11}. Wolfram \cite{kn:Wolf02} exhibited glider collisions with long periods of after-development and several filters for detecting gliders and defects, and Martin \cite{kn:Mar00} designed an algebraic group of order four to represent collisions between basic gliders. A number of new glider guns, self-organization by structures, collisions, and glider-based logic gates were reported in \cite{kn:MAM06}. Guan \cite{kn:Guan12} develops a description of rule 54 dynamics with Bernoulli shift and symbolic sequences. Redeker \cite{kn:Red10} discusses how a flexible time can be represented in evolutions of rule 54.  An exhaustive analysis about solitons in rule 54 was presented in \cite{kn:MAC12}, and a projection of rule 54 affected with memory was studied in \cite{kn:MAA13}. Initial analysis of glider representation with rule 54 by de Bruijn and cycles diagrams was given in \cite{kn:MAM08} (see also \cite{kn:repR54}).

\section{Rule 54}

A one-dimensional cellular automaton (CA) is represented by an infinite array of {\it cells} $x_i$ where $i \in \mathbb{Z}$ and each $x$ takes a value from a finite alphabet $\Sigma$. Thus, a sequence of cells \{$x_i$\} of finite length $n$ represents a string or {\it global configuration} $c$ on $\Sigma$. The set of finite configurations, represented as $\Sigma^n$, is denoted by $\Phi$. The CA evolution is given by a sequence of configurations $\{c_i\}$ on $\Phi$:

\begin{equation}
\Phi(c^t) \rightarrow c^{t+1},
\label{globalFunction}
\end{equation}

\noindent where $t$ is time and every global state of $c$ is defined by a sequence of cells. Also the cells of each configuration $c^t$ are updated at the next configuration $c^{t+1}$ simultaneously by a local function $\varphi$ as follows:

\begin{equation}
\varphi(\ldots, x_{i-1}^t, x_{i}^t, x_{i+1}^t, \ldots) \rightarrow x_i^{t+1}.
\label{localFunction}
\end{equation}

A one-dimensional CA can be described by two parameters $(k,r)$ \cite{kn:Wolf94}. Where $k = |\Sigma|$ is a number of states and $r$ is a neighborhood radius. ECAs are defined by parameters $(k=2,r=1)$. 

In all constructs described in the paper, we apply  {\it periodic boundary conditions} to obtain finite configurations of $\Phi$ by concatenating the first cell with the last one to form a ring.

The local transition function $\varphi$ of ECA rule 54 follows:

\[
\varphi_{R54} = \left\{
	\begin{array}{lcl}
		1 & \mbox{if} & 101, 100, 010, 001 \\
		0 & \mbox{if} & 111, 110, 011, 000
	\end{array} \right. .
\]

\begin{figure}[th]
\begin{center}
\subfigure[]{\scalebox{0.35}{\includegraphics{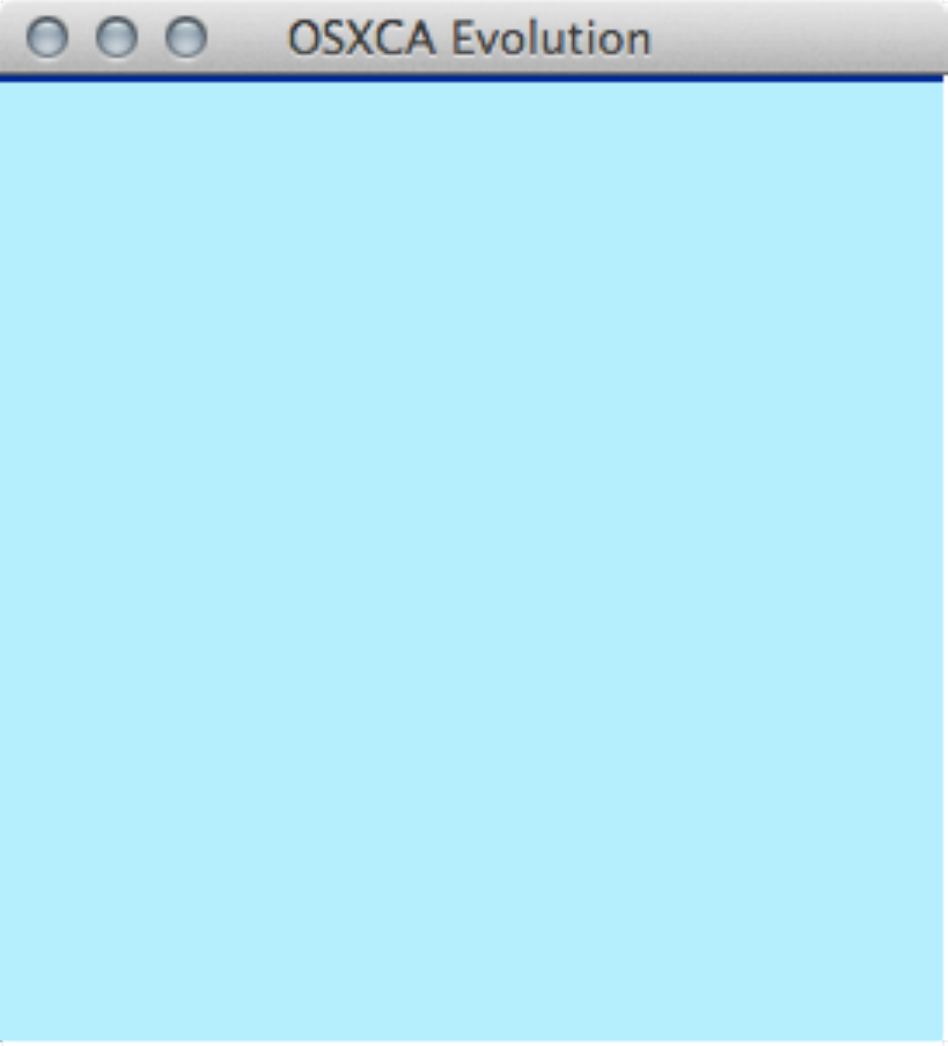}}} \hspace{0.1cm}
\subfigure[]{\scalebox{0.35}{\includegraphics{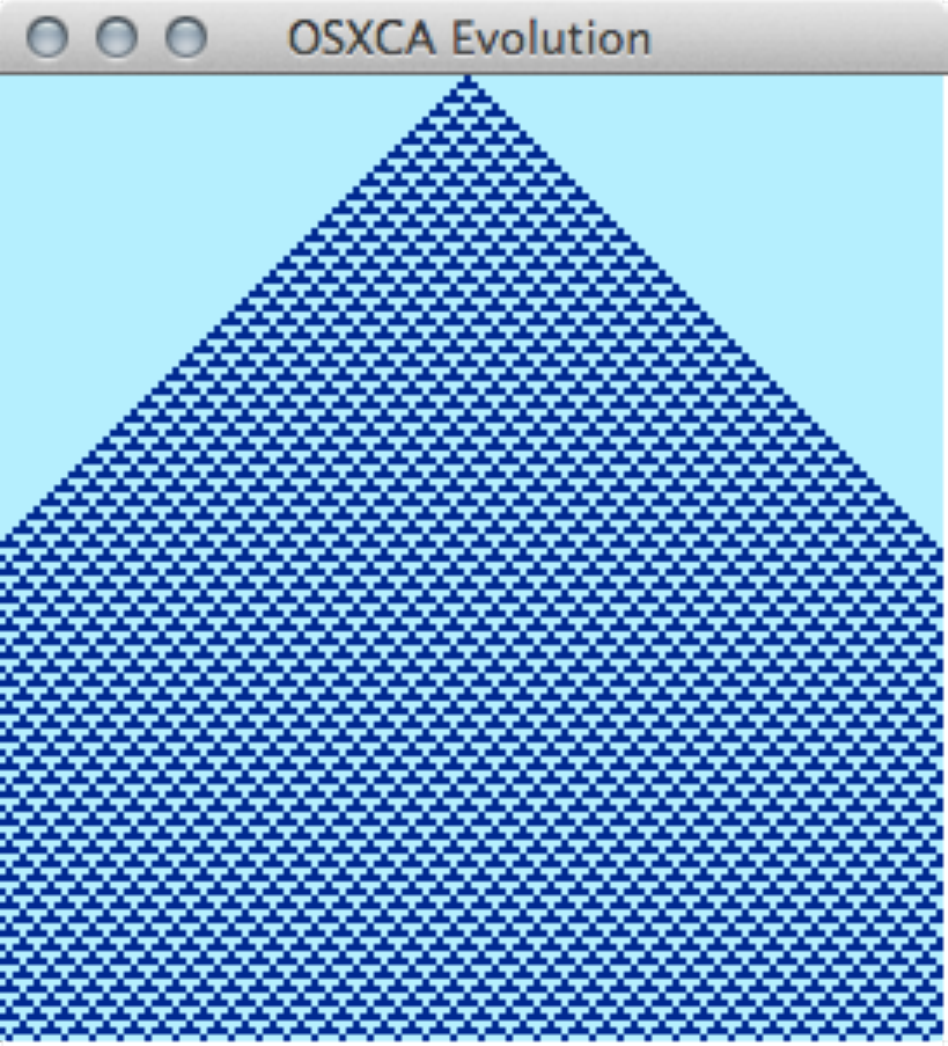}}} \hspace{0.1cm}
\subfigure[]{\scalebox{0.35}{\includegraphics{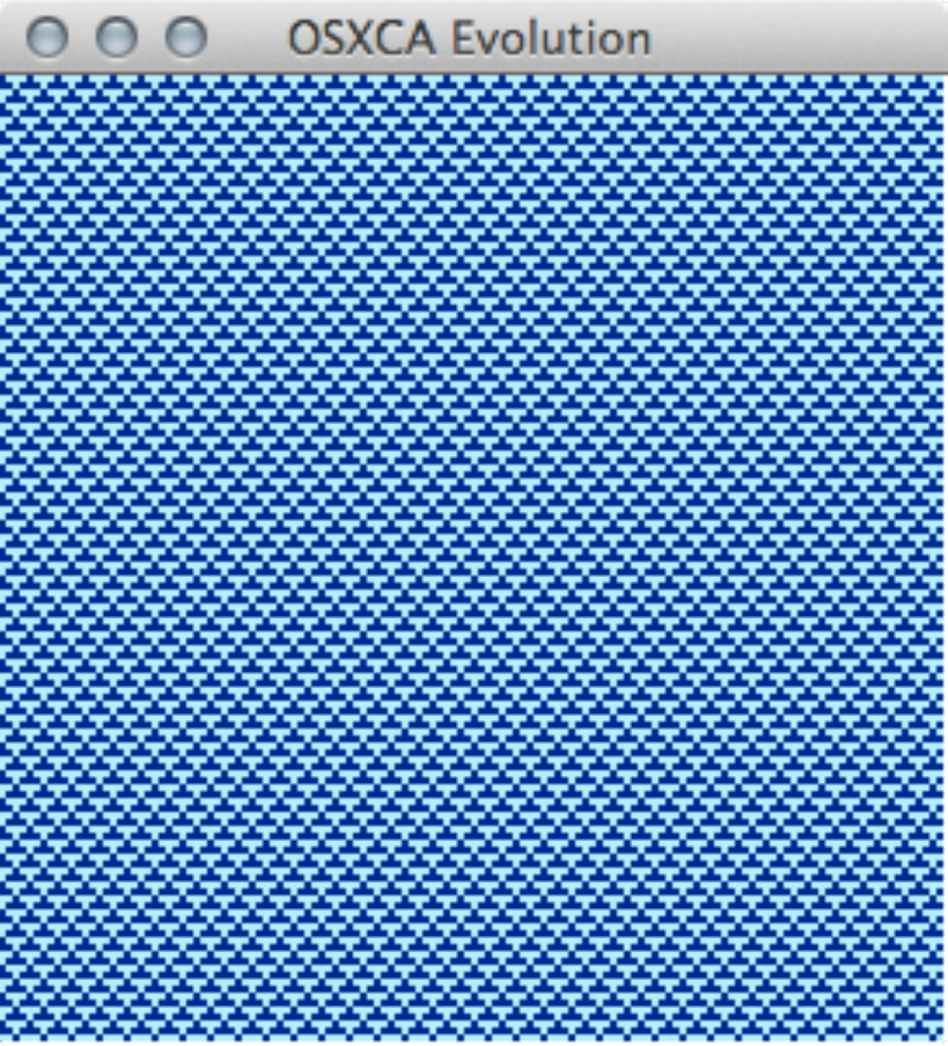}}} \hspace{0.2cm}
\subfigure[]{\scalebox{0.29}{\includegraphics{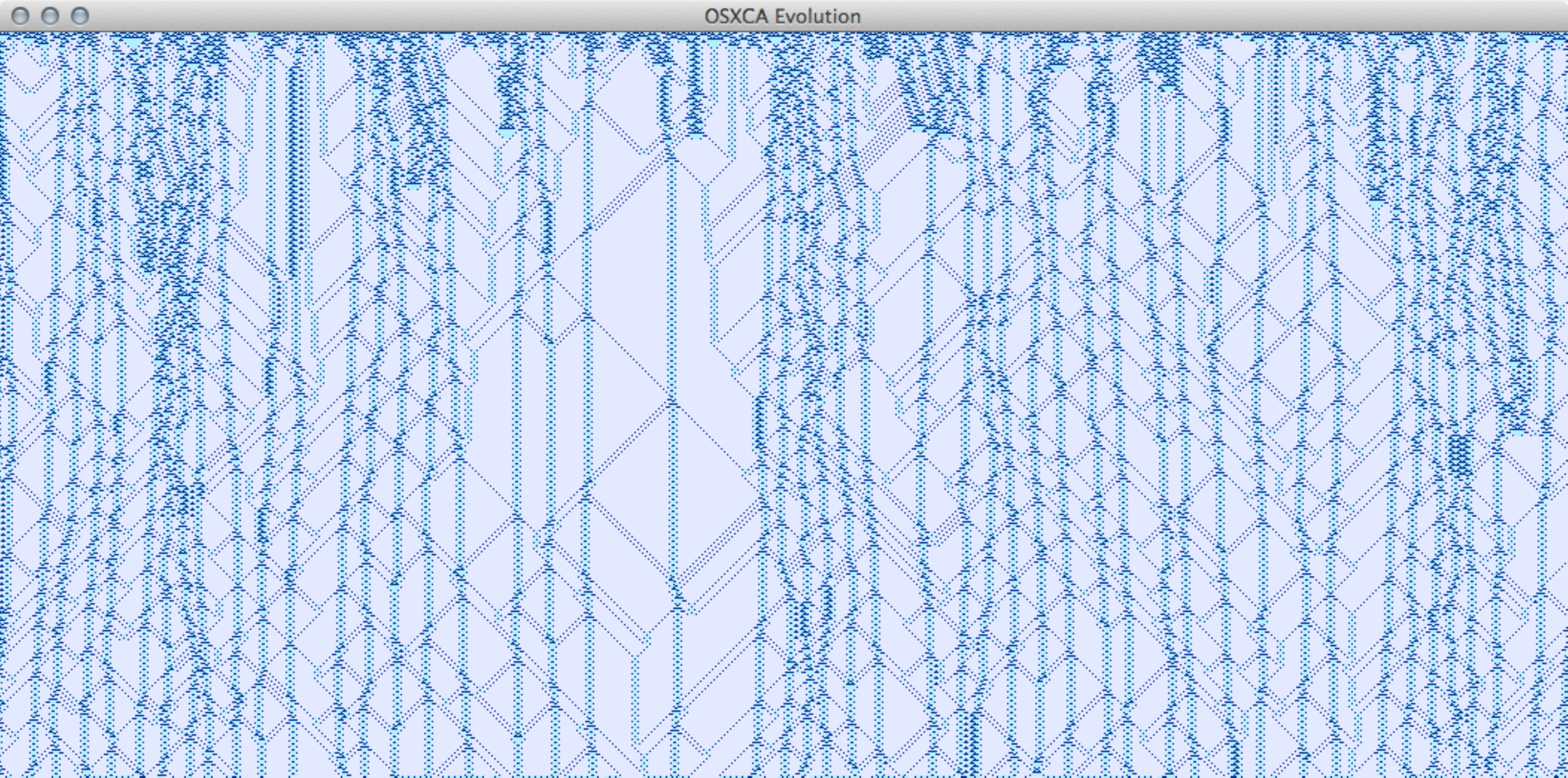}}}
\end{center}
\caption{Exemplar scenarios of space-tiime evolution in rule 54 with 100 cells for 100 generations. (a) A single state of $\Sigma$ dominates the initial condition. (b) A single cell is in state 1; all other cells are in state 0. (c) Periodic background. (d) Random initial condition with an initial density of 50\% on 1000 cells for 500 generations (a filter is applied).}
\label{dynamicR54}
\end{figure}

The binary sequence $00110110$ represents rule number 54 in decimal notation. Initially $\varphi_{R54}$ presents an initial probability of 50\% to each state, and thus the frequency to appear is the same.

Figure~\ref{dynamicR54} displays some typical snapshots with rule 54. We have chosen classic or specific initial conditions to order capture different behaviors. Indeed, this set of figures can represent several CA classes: (a) could represent class I with a uniform evolution, (b) and (c) class II with periodic evolutions, and (d) class IV with complex dynamics.

Rule 54 is a discrete analog of an active, nonlinear one-dimensional medium. Assume each cell is a micro-volume, which takes two states:  resting (0) and excited (1). When a single micro-volume is perturbed, its corresponding cell takes state 1. The perturbation/activation spreads to neighbors of the initially excited micro-volume: $\{ 100, 001, 101 \} \rightarrow 1$. For example, the transition $100 \rightarrow 1$ encodes the following activation mechanism: if the left neighbor of a resting cell is excited, the resting cell excites. Transition $000 \rightarrow 0$ indicates the simple fact that the medium could not activate itself; that is, excitation cannot develop from a totally resting medium.

The three most interesting transitions are $\{ 111, 110, 011 \} \rightarrow 1$. They encode the following fact: if an excited micro-volume has at least one excited neighboring micro-volume, then this micro-volume returns to a resting state. This can be interpreted as a mutual inhibition. Each excited micro-volume inhibits excitation of its excited neighboring micro-volume. These features of rule 54 make it also interesting from neurophysiology and machine vision (one-dimensional artificial retina) points of view.

\section{Representation of Gliders in Rule 54}

This section discusses approaches toward a description of gliders in rule 54. These approaches use tiling theory \cite{kn:GS82}, de Bruijn diagrams \cite{kn:Mc91, kn:Mc09}, and cycle diagrams \cite{kn:WL92}.

\subsection{Tiles in Rule 54}

Gliders in rule 54 can be represented by polygons as rhomboids. The periodic background, called ether, is represented in rule 110 for a family of triangles \cite{kn:Mc99, kn:MMS06}, although with some differences from rule 54.

\begin{figure}[th]
\centerline{\includegraphics[width=4.2in]{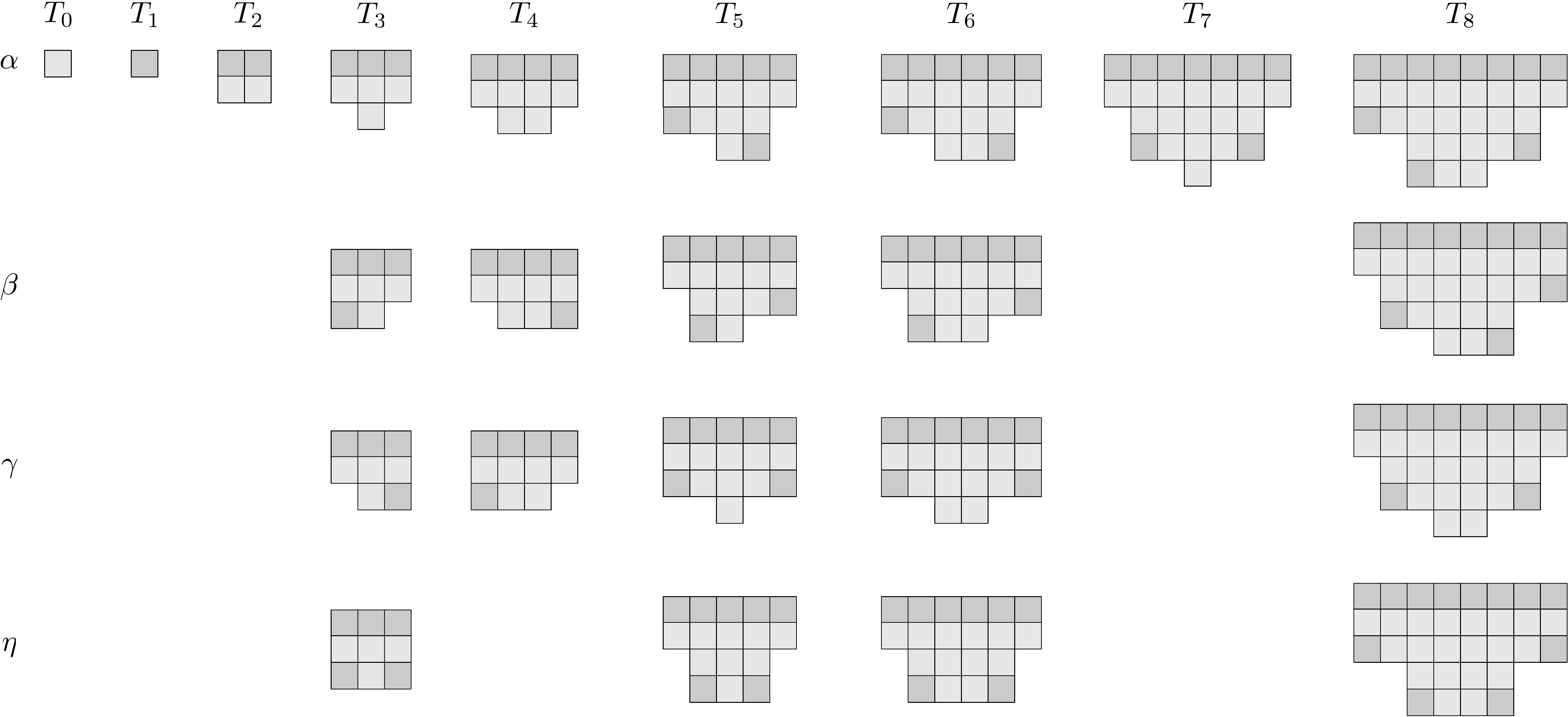}}
\caption{Examples of tiles $\mathcal T_{R54}$ derived from space-time configurations in rule 54.}
\label{tilesR54}
\end{figure}

A {\it plane of tiles} $\mathcal T$ is a countable family of closed sets $\mathcal T = \{T_{0},T_{1},\ldots\}$ covering the plane without intervals or intersections \cite{kn:GS82} (the ``plane'' is the Euclidian plane $\mathbb Z \times \mathbb Z$ in elementary geometry). Therefore, this can be defined as a join of sets (called a mosaic $\mathcal T$):

\begin{equation}
\mathcal T = \bigcup_{i=0}^{n}{T_{i}} \mbox{ $\forall$ } n \in \mathbb Z^+_0;
\end{equation}

\noindent consequently, every set is disjoint $T_i \cap T_j$. Thus, the set of tiles for rule 54 is represented as $\mathcal T_{R54}$. Figure~\ref{tilesR54} displays a number of mosaics of $\mathcal T_{R54}$.

Table~\ref{tilesRelationR54} shows relations between tiles in rule 54. A row represents the tile type and a column represents the size of a tile. There are a limited number of kinds but an infinite number of sizes. 

\begin{figure}[th]
\centerline{\includegraphics[width=2.5in]{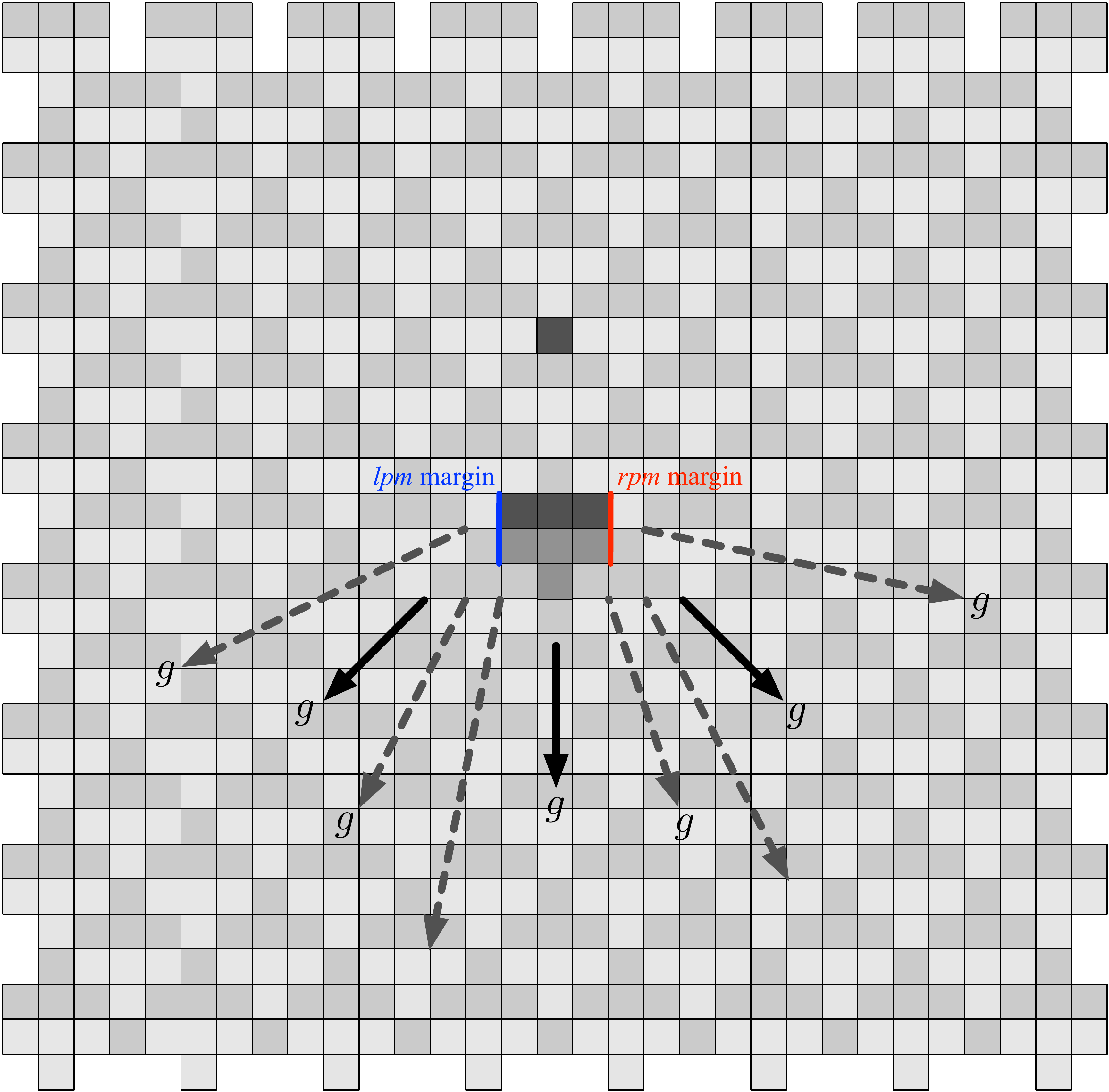}}
\caption{Periodic background in rule 54 is composed with two tiles of the set $\mathcal T_{R54}$: $T_1$ and $T_3^{\alpha}$.}
\label{tilesB}
\end{figure}

Rule 54 can be studied as a tiling problem (as was proposed in rule 110 by McIntosh \cite{kn:Mc99, kn:Mc00}). Figure~\ref{tilesR54} shows the relation of $\mathcal T_{R54}$ for the first nine polygons, which is summarized in Table~\ref{tilesRelationR54}. If one relation is missing, this means that such a polygon cannot be constructed for rule 54.

\begin{table}[th]
\centering
\begin{tabular}{c|cccccccccc}
$T$ & 0 & 1 & 2 & 3 & 4 & 5 & 6 & 7 & 8 & $\cdots$ \\
\hline
$\alpha$ & \checkmark & \checkmark & \checkmark & \checkmark & \checkmark & \checkmark & \checkmark & \checkmark & \checkmark & $\cdots$ \\
$\beta$ & & & & \checkmark & \checkmark & \checkmark & \checkmark & & \checkmark & $\cdots$ \\
$\gamma$ & & & & \checkmark & \checkmark & \checkmark & \checkmark & & \checkmark & $\cdots$ \\
$\eta$ & & & & & \checkmark & \checkmark & \checkmark & & \checkmark  & $\cdots$
\end{tabular}
\caption{Relation of tiles $\mathcal T_{R54}$ in rule 54.}
\label{tilesRelationR54}
\end{table}

Figure~\ref{tilesB} displays the composition for the periodic background in rule 54. It is composed of two tiles: $T_1$ and $T_3^{\alpha}$. Arrows indicate possible directions in which a glider could emerge. Solid arrows display known gliders and dotted arrows display possible unknown gliders.

\subsection{Gliders in Rule 54}

A {\it glider} is a compact group of non-quiescent states traveling along the CA lattice. To represent gliders in rule 54, we follow the notation of Boccara et al.~\cite{kn:BNR91}.

\begin{figure}[th]
\begin{center}
\subfigure[]{\scalebox{0.18}{\includegraphics{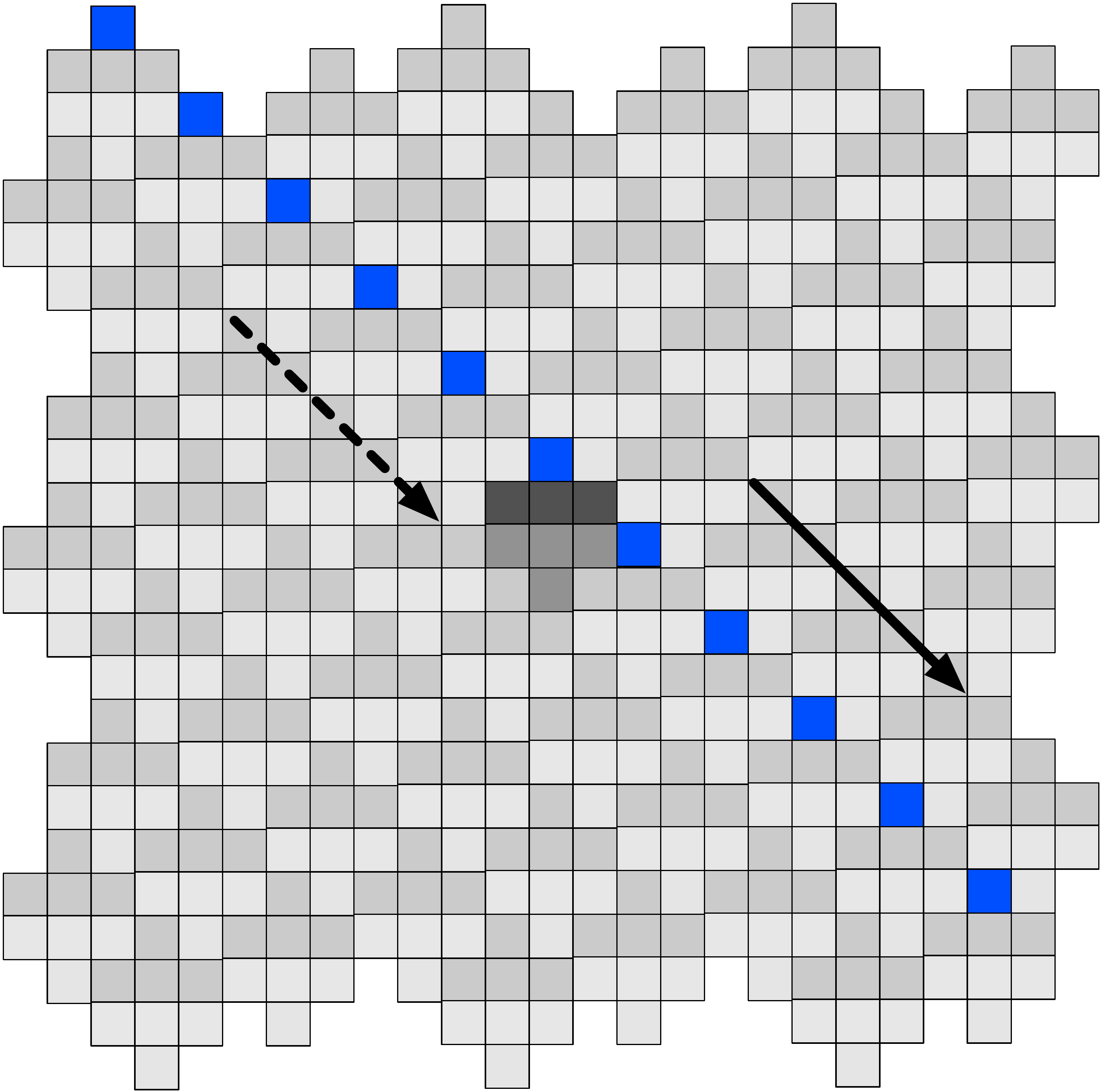}}} \hspace{0.1cm}
\subfigure[]{\scalebox{0.18}{\includegraphics{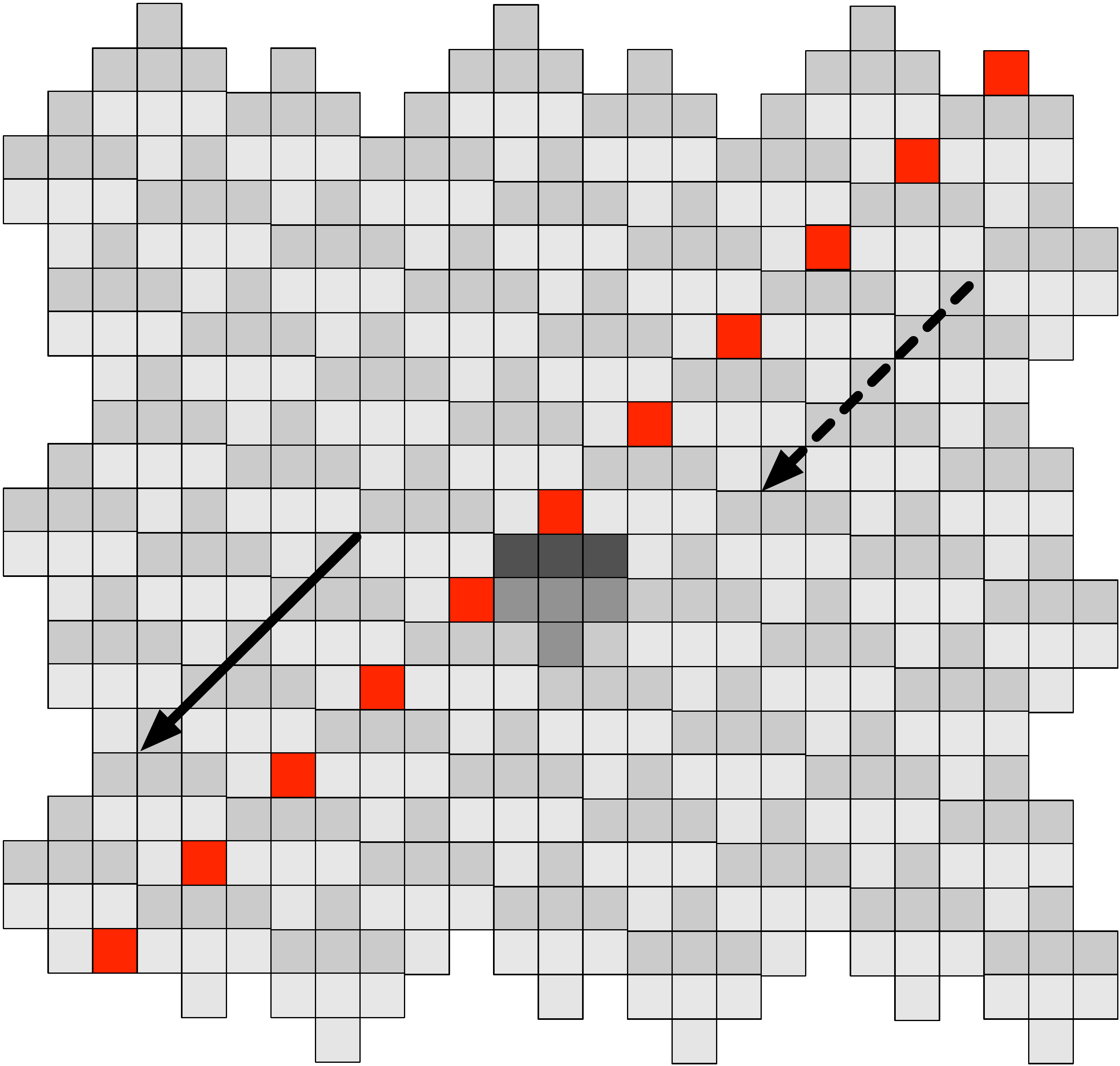}}} 
\end{center}
\caption{Tiles description for primitive gliders in rule 54.}
\label{tilesB2}
\end{figure}

\begin{figure}[th]
\begin{center}
\subfigure[]{\scalebox{0.145}{\includegraphics{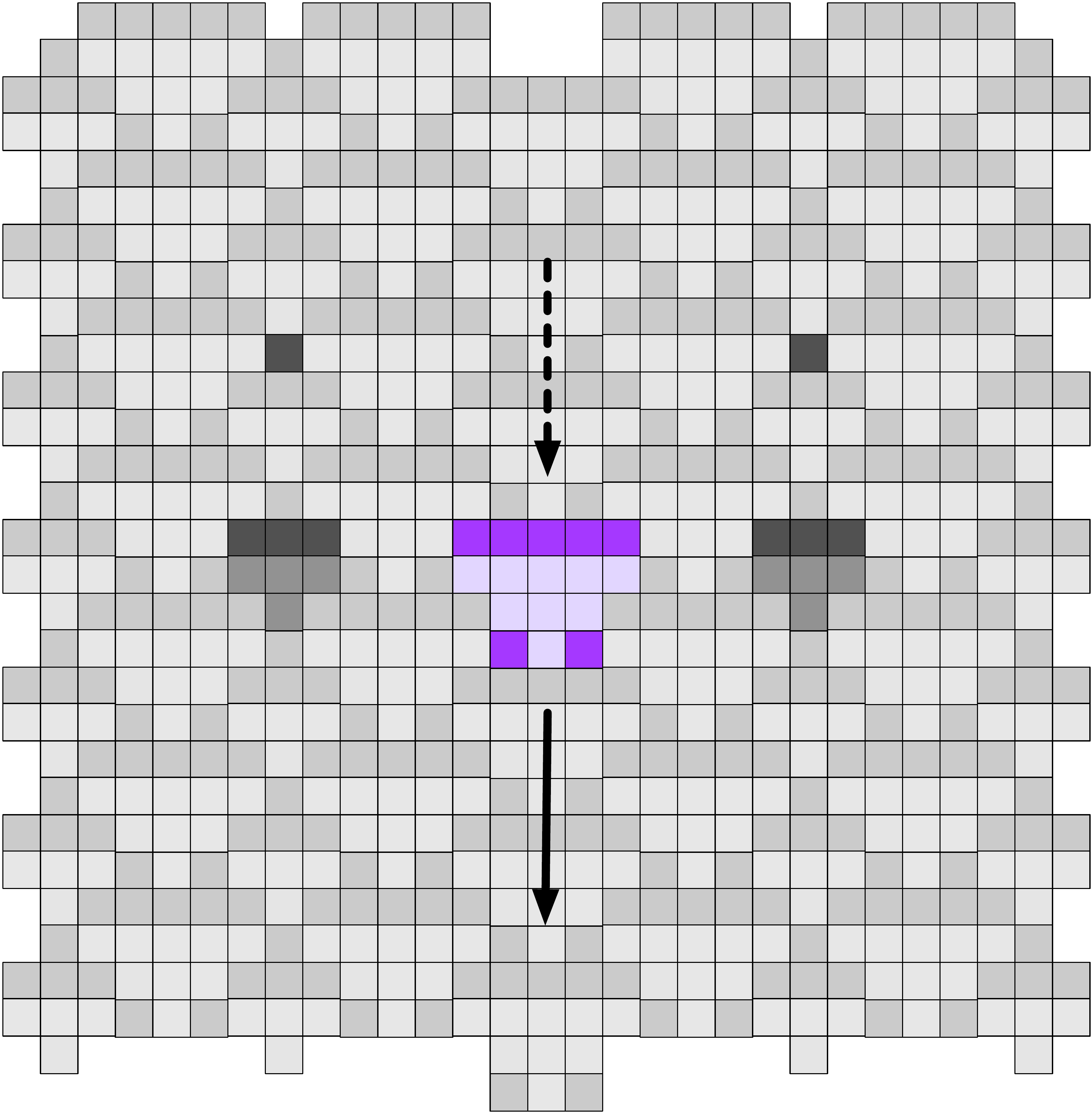}}} 
\subfigure[]{\scalebox{0.145}{\includegraphics{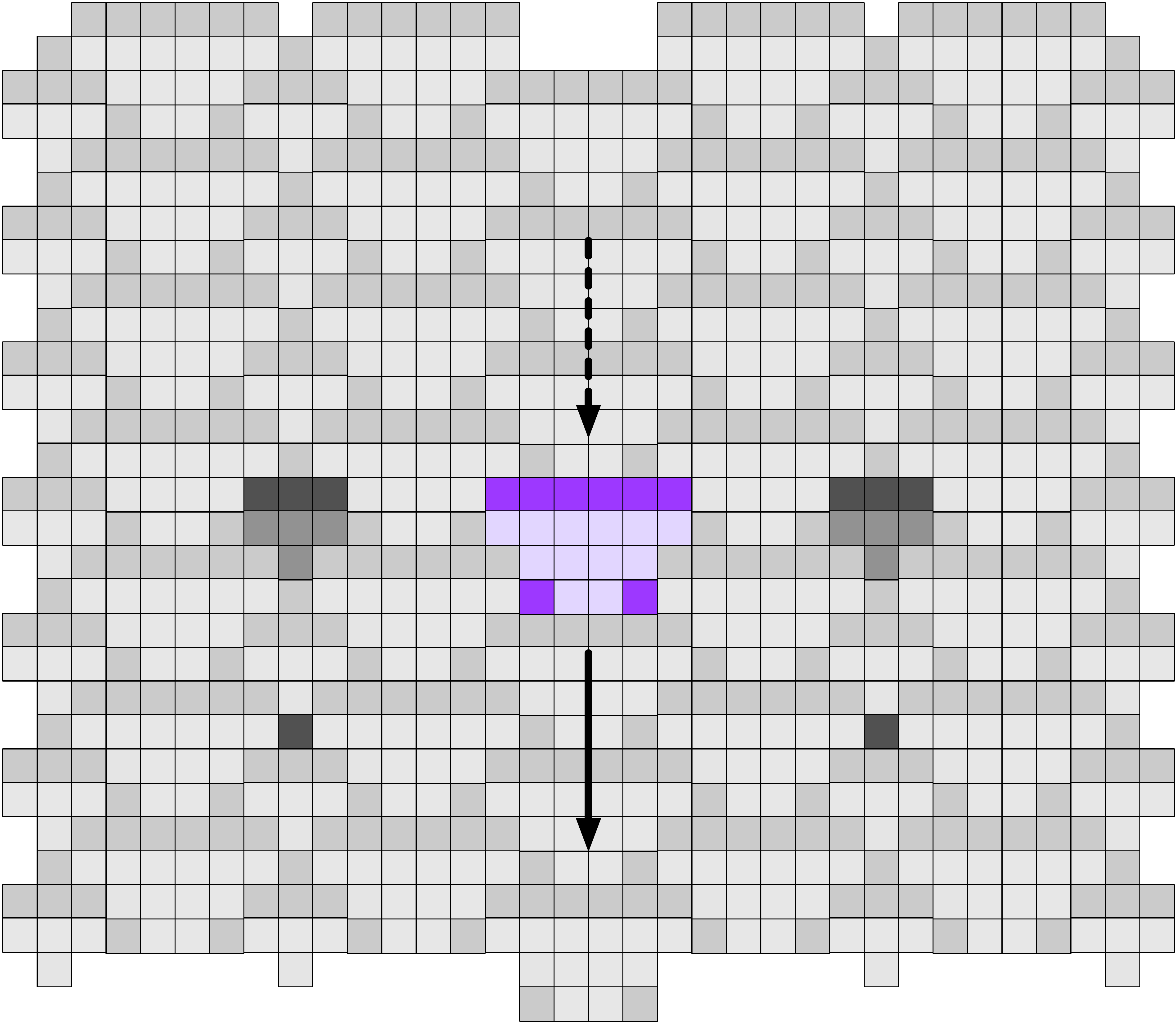}}} 
\end{center}
\caption{Tiles description of composed gliders in rule 54. Both gliders are still life configurations because they are stationary.}
\label{tilesB3}
\end{figure}

Rule 54 has two identical primitive gliders traveling in opposite directions: $\overrightarrow{w}$ glider (Figure~\ref{tilesB2}a) and $\overleftarrow{w}$ glider (Figure~\ref{tilesB2}b) traveling with the speed of light, that is, translating one cell per iteration. Two stationary gliders can be interpreted as still life configurations in one dimension. They are gliders $g_o$ (Figure~\ref{tilesB3}(a)) and $g_e$ (Figure~\ref{tilesB3}(b)).

\begin{figure}[th]
\centerline{\includegraphics[width=4.13in]{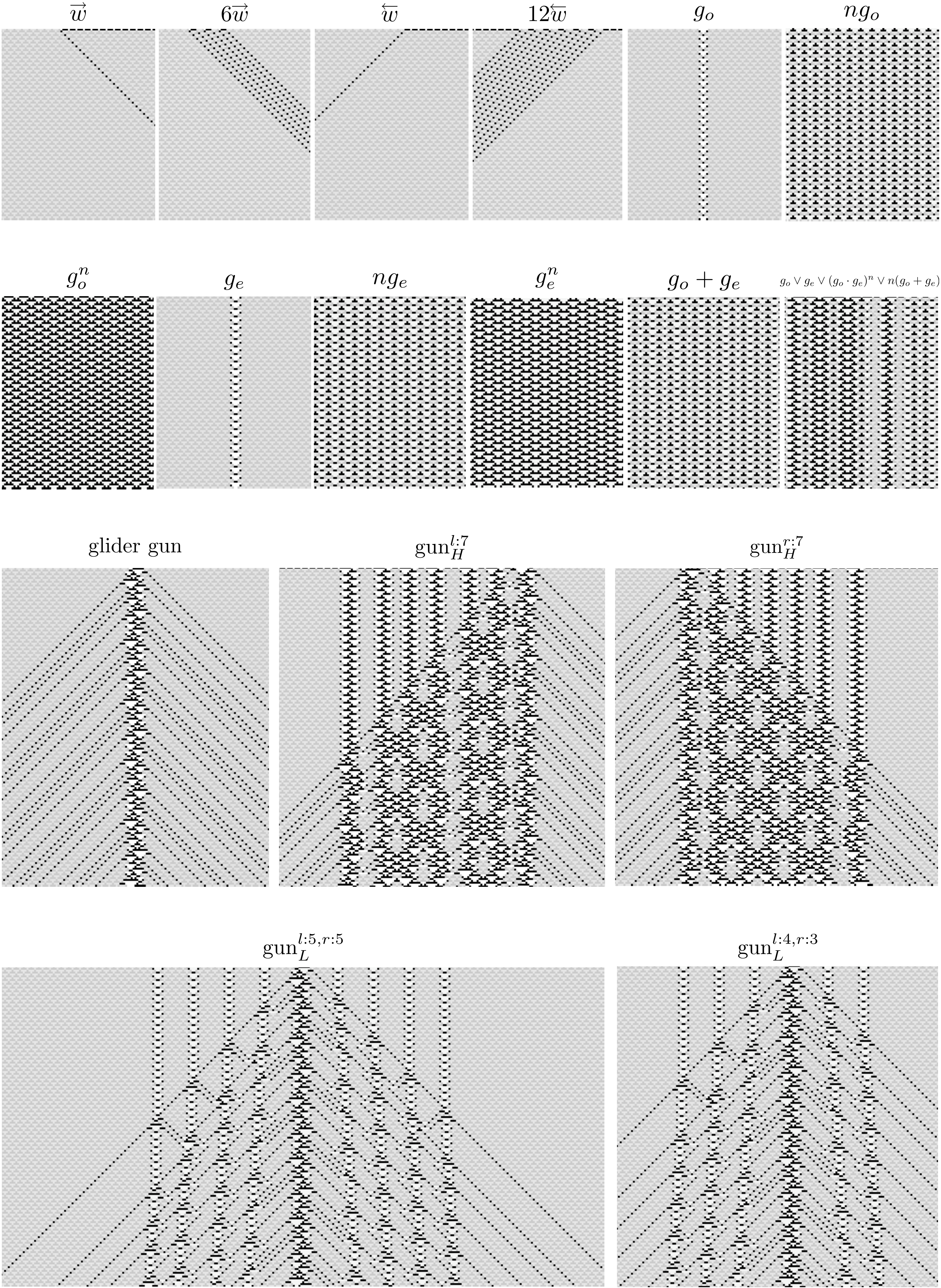}}
\caption{Classification of gliders in rule 54. We illustrate every glider and packages, extensions, and compositions of them.}
\label{glidersR54}
\end{figure}

This way, we can display each glider, enumerating their properties. Figure~\ref{glidersR54} gives a systematic representation of gliders in rule 54, and Table~\ref{tablaGlidersR54} summarizes most basic properties.

Let $e_1$ and $e_2$ represent glider phases in the periodic background. Thus we have four gliders: $\overrightarrow{w}, \overleftarrow{w}, g_{o}, g_{e}$; and a compound glider: the glider gun. The speed $v_g$ of a glider is evaluated using the period between displacements. Column {\it Cap} in Table~\ref{tablaGlidersR54} shows if a glider is able to cover the full space without gaps.

\begin{table}[th]
\centering
\begin{tabular}{|c|c|c|c|}
\hline
Structure & $v_{g}$ & Lineal Volume & Cap \\
\hline \hline
$e_{1}$ & 2/2 = 1 & 4 & T \\
\hline
$e_{2}$ & 2/2 = 1 & 4 & T  \\
\hline
$\overrightarrow{w}$ & 2/2 = 1 & 2 & P \\
\hline
$\overleftarrow{w}$ & -2/2 = -1 & 0-4 & P \\
\hline
$g_{o}$ & 0/4 = 0 & 6-2 & T \\
\hline
$g_{e}$ & 0/4 = 0 & 7-3 & T \\
\hline
glider gun & 0/32 = 0 & 14-4 & P \\
\hline
\end{tabular}
\caption{Properties of gliders in rule 54.}
\label{tablaGlidersR54}
\end{table}

Rule 54 exhibits a relatively small number of gliders, which makes it particularly attractive for discretization and formal representation. We can obtain an exact representation of gliders in rule 54 and show how to construct specific initial conditions based on glider phases. A {\it phase} means a unique string that represents the glider in the initial condition. Therefore, a finite number of different strings represent the set of valid strings where a glider can be initialized \cite{kn:MMS08}.

\subsection{De Bruijn diagram}

For a one-dimensional CA of order $(k,r)$ and a finite alphabet given $\Sigma$, its de Bruijn diagram is defined as a directed graph with $k^{2r}$ vertices and $k^{2r+1}$ edges. Vertices are labeled with elements of the alphabet of length $2r$, that is, neighborhood states. An edge is directed from vertex $i$ to vertex $j$ if and only if the $2r-1$ final symbols of $i$ are the same as $2r-1$ initial symbols in $j$, forming a neighborhood of $2r+1$ states represented by $i \diamond j$. In this case, the edge connecting $i$ to $j$ is labeled by $\varphi(i \diamond j)$ (the value of the neighborhood defined by the local function) \cite{kn:Voor96, kn:Voor08}.

Thus, the de Bruijn diagram can be constructed as follows:
 
\begin{equation}
	M_{i,j} = \left\{\begin{array}{ll}
			        1& \mbox{if } j = ki, ki+1, \ldots, ki+k-1 \mbox{ (mod } k^{2r}) \\
		           	 0 & \mbox{in other case} \\
		       \end{array}
			\right.
\label{eq-Bruijn}
\end{equation}

\begin{figure}[th]
\centerline{\includegraphics[width=4in]{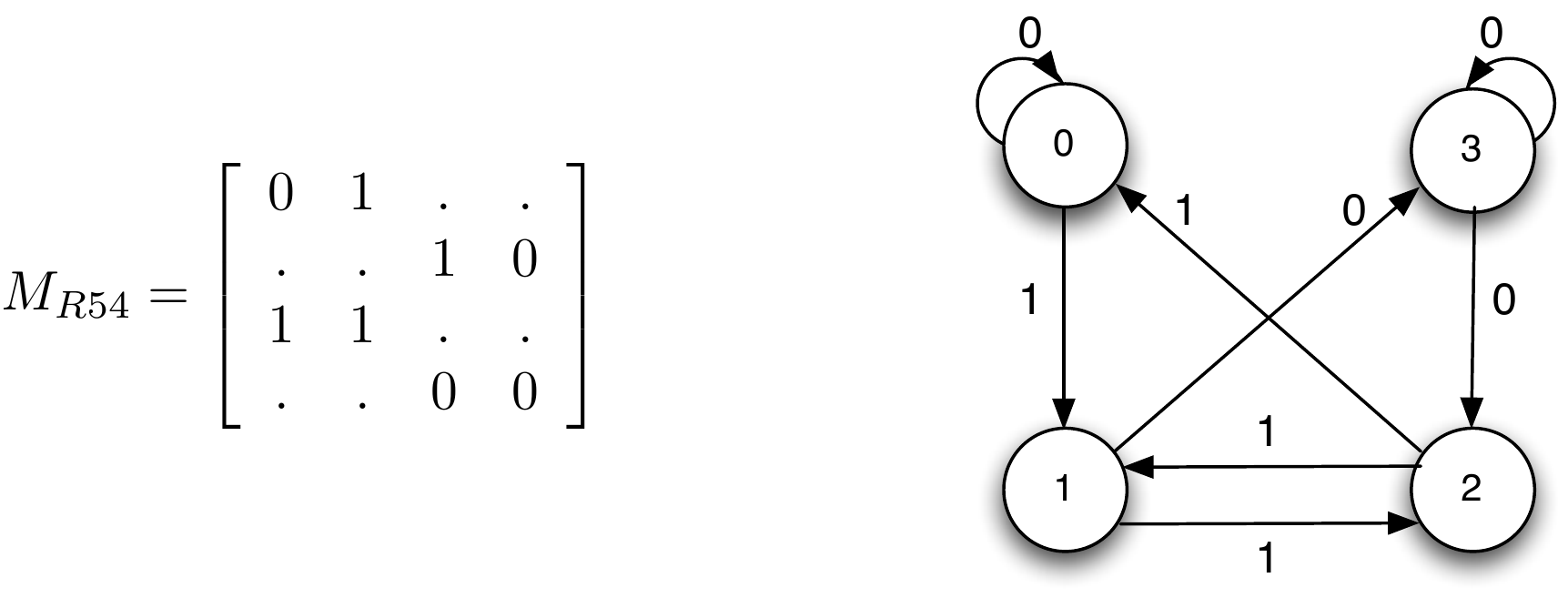}}
\caption{de Bruijn diagram of rule 54.}
\label{deBruijnR54}
\end{figure}

Module $k^{2r}=2^{2}=4$ represents the number of vertices in the de Bruijn diagram, and $j$ takes values from $k*i=2i$ to $(k*i)+k-1=(2*i)+2-1=2i+1$. The vertices (indexes of $M$) are labeled by fractions of neighborhoods beginning with 00, 01, 10, and 11; the overlap determines each connection. Figure~\ref{deBruijnR54} displays rule 54's matrix evolution and its de Bruijn diagram.

Paths in the de Bruijn diagram may represent chains, configurations, or classes of configurations in the evolution space. Also, fragments of the diagram itself are useful in discovering periodic blocks of strings, pre-images, codes, and cycles~\cite{kn:Mc09, kn:Voor96}.

After the de Bruijn diagram is completed, we can calculate an extended de Bruijn diagram~\cite{kn:Mc09}. An extended de Bruijn diagram takes into account more significant overlapping of neighborhoods. Thus, we represent $M^{(2)}_{R54}$ by indexes $i=j=2r*n$, where $n \in \mathbb Z^{+}$. The de Bruijn diagram grows exponentially, order $k^{2r^{n}}$, for each $M^{(n)}_{R54}$; the basic de Bruijn diagram is obtained for $n=1$ (Figure~\ref{deBruijnR54}).

An important indication derived from de Bruijn diagrams is that the set of regular expressions $\Psi_{R54}$ describing all possible strings to initialize gliders in rule 54. Of course, this representation does not include codes to initialize packages or groups of gliders. Therefore, the number of sequences $w$ in the set $\Psi_{R54}$ is the union of the periods for every glider, as follows:

\begin{equation}
\Psi_{R54} = \bigcup_{i=1}^{p}{w_{i,g}} \mbox{ } \forall \mbox{ } (w_i \in \Sigma^* \wedge g \in \cal G),
\end{equation}

\noindent where $\cal G$ is the whole set of gliders in rule 54 and $p > 0$ its period. This way, we can speak of a regular language $L_{R54}$ that is constructed from the expressions of $\Psi_{R54}$. We  notice that this language is a subset of the whole language in rule 54, because it is defined by regular expressions derived from gliders. Therefore, the regular language $L_{R54}$ is defined as follows:

\begin{equation}
\small
L_{R54} = \{w | w \mbox{ operating under the basic rules: } \cdot,+,* \mbox{ from } \Psi_{R54}\}.
\end{equation}

Let us calculate de Bruijn diagrams for gliders $\overrightarrow{w}$ and $\overleftarrow{w}$ with periodic background. Table~\ref{tablaGlidersR54} shows that these gliders move two cells in a time. Then the extended de Bruijn diagram of order $M^{(2)}_{R54}$ would be necessary to extract a cyclic structure of gliders (all extended de Bruijn diagrams are calculated with NXLCAU21, a free software developed by H. V. McIntosh \cite{kn:McNX}). These diagrams  can show all possible relations, but cycles are important for us for detect gliders or other periodic patterns.

\begin{figure}[th]
\centerline{\includegraphics[width=4.2in]{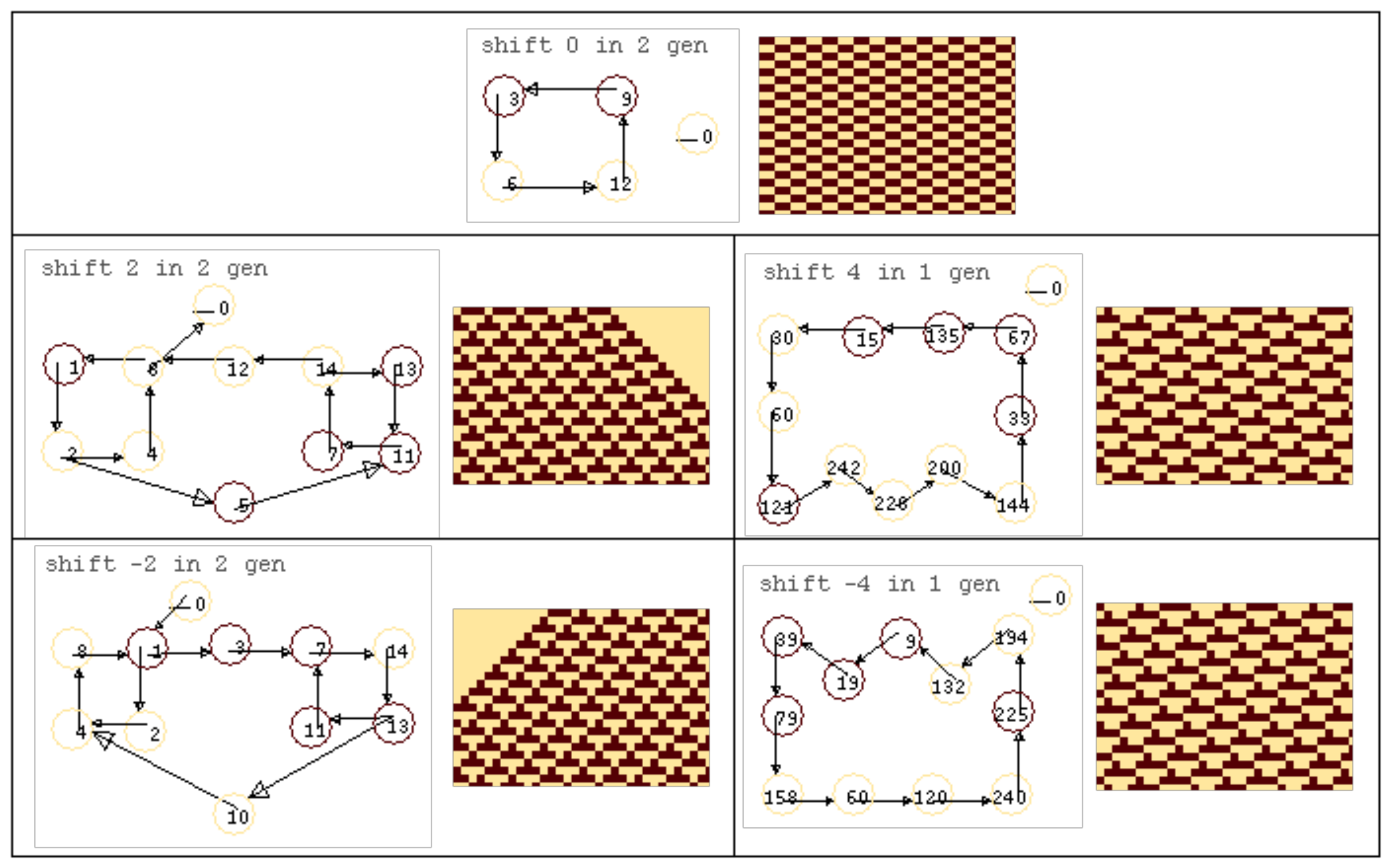}}
\caption{De Bruijn diagrams determining primitive gliders, periodic background, and other meshes (0,2), (4,1), (-4,1) in rule 54.}
\label{deBruijn-1}
\end{figure}

Figure~\ref{deBruijn-1} displays de Bruijn diagrams with shift registers to the right $(+)$ or to the left $(-)$. A glider can be identified as a cycle and the glider's interaction will be a connection with other cycles. Diagram (2,2) ($x$-displacements, $y$-generations), displays periodic strings moving two cells to the right in two steps, that is, the period between displacement in the periodic background of a $\overrightarrow{w}$ glider. This way, we can enumerate each string for every structure in this domain.

\begin{itemize}
\item Periodic background $e$ is fixed as:

\begin{itemize}
\item vertices $(1, 2, 4, 6)$ $\equiv$ $e_1 = 1000$
\item vertices $(13, 11, 7, 14)$ $\equiv$ $e_2 = 1110$
\end{itemize}

\item $\overrightarrow{w}$ glider is placed as:

\begin{itemize}
\item vertices $(1, 2)$ $\equiv$ $\overrightarrow{w}_1 = 10$
\item vertices $(12, 6)$ $\equiv$ $\overrightarrow{w}_2 = 00$
\end{itemize}

\item $\overleftarrow{w}$ glider is placed as:

\begin{itemize}
\item vertices $(1, 3, 7, 14)$ $\equiv$ $\overleftarrow{w}_1 = 1110$
\item vertices $(13, 10, 4, 8)$ $\equiv$ $\overleftarrow{w}_2 = 1000$
\end{itemize}
\end{itemize}

The periodic background in phase one represents the string 1000 and  phase two the string 1110. Also, this diagram has a positive orientation of cycles and shows that relations of vertices $(1, 2, 4, 6)$ and $(13, 11, 7, 14)$ that represent all possible phases where a $\overrightarrow{w}$ glider may be initialized. However, the existence of this glider is related to both cycles.

Diagrams $(0,2)$, $(4,1)$, and $(-4,1)$ display three different periodic backgrounds that cannot coexist with gliders but can cover the whole evolution space.

Rule 54 has a particular characteristic because the periodic background needs a displacement to preserve the existence of gliders. Figure~\ref{deBruijn-1} shows four cycles, three of them self-contained and one  starts with stable state. Evolution fragments in the same picture show what kinds of gliders are defined by these cycles. For example, we can see a large cycle following the vertices $(1, 2, 5, 11, 13, 14, 12, 6)$. This cycle is  equivalent to the periodic string $10111000$, which produces an evolution space covered with just a pair of $\overrightarrow{w}$ gliders. Finally, a fourth cycle, represented by the cycle 0, determines a transition between two different patterns, known as a ``fuse configurations'' \cite{kn:Mc09}. The periodic background is formed by a cycle of length four, and the existence of gliders is determined by other cycles. Therefore, we can see that the problem of representing gliders by de Bruijn diagrams is reduced to the classification of cycles.

To represent gliders $g_o$ and $g_e$, we should construct de Bruijn diagrams of order $M^{(4)}_{R54}$, because gliders have period 4 (see Table~\ref{tablaGlidersR54}). These gliders can be considered as still life configurations because they are stationary structures.

\begin{figure}[th]
\centerline{\includegraphics[width=4.2in]{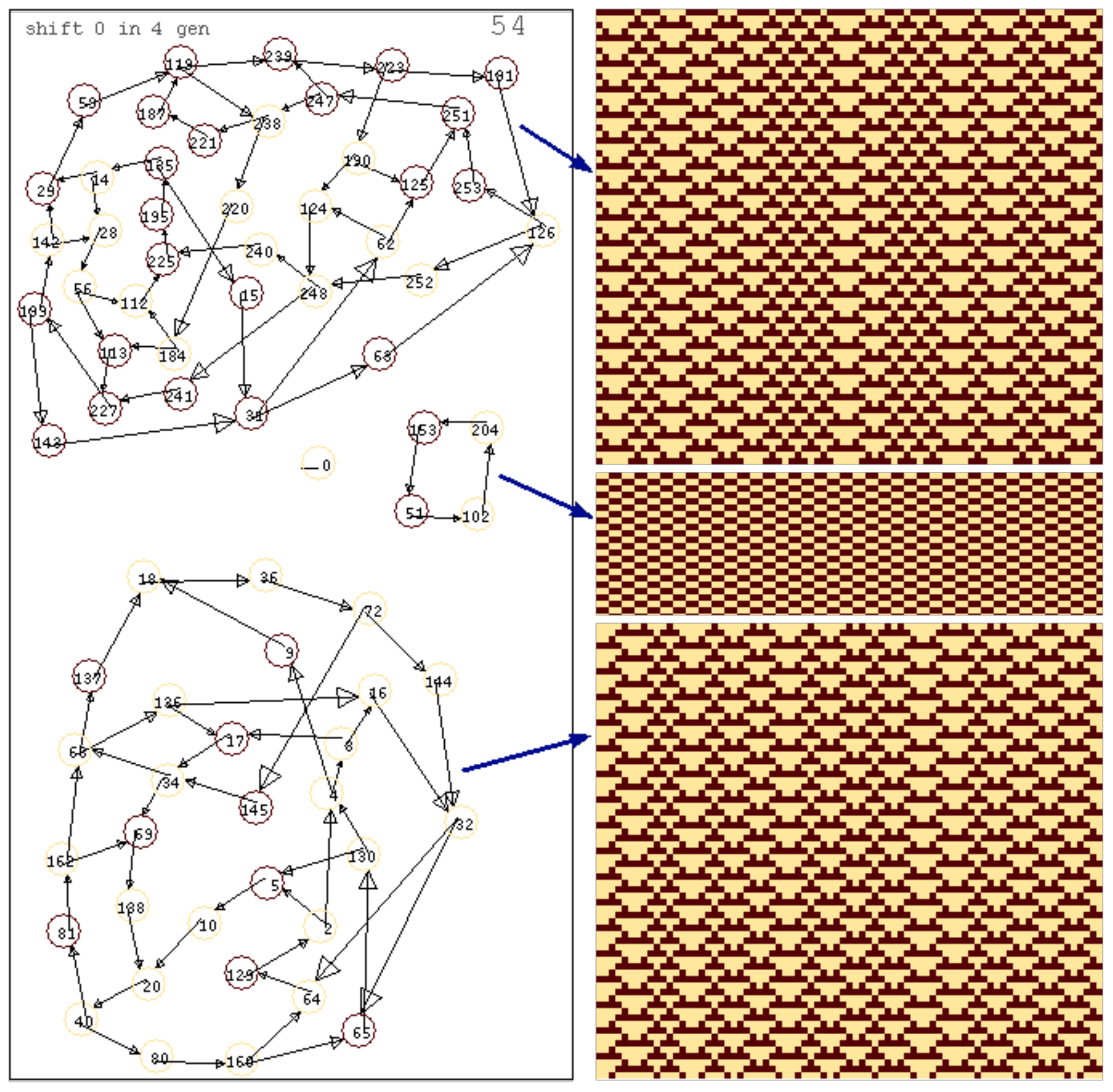}}
\caption{De Bruijn diagram representing stationary gliders.}
\label{deBruijn-2}
\end{figure}

Figure~\ref{deBruijn-2} shows the full de Bruijn diagram $(0,4)$ used to calculate $g_o$ and $g_e$ gliders. There are four main cycles: two largest cycles represent phases of $g_o$ and $g_e$ plus its periodic background; and two smaller cycles characterize two different periodic patterns in rule 54 including the stable state.

Again, to extract phases we shall follow routes in the diagram and enumerate all the routes, that is, their regular expressions. The larger cycles contain internal cycles that represent each glider phase. So the periodic background is represented by two cycles and they relate all possible phases for $g_o$ and $g_e$ gliders.

The left cycle in the diagram of Figure~\ref{deBruijn-2} represents the whole phases of gliders $g_o$ and $g_e$ with the periodic background $e_1$, vertices $(17, 34, 68, 136)$, and the right cycle represents phases with the periodic background $e_2$, vertices $(221, 187, 119, 238)$. Therefore, we can extract periodic sequences to encode gliders traveling alone or in trains of gliders. Encoding samples are provided with some strings. 

\begin{figure}[th]
\begin{center}
\subfigure[]{\scalebox{0.4}{\includegraphics{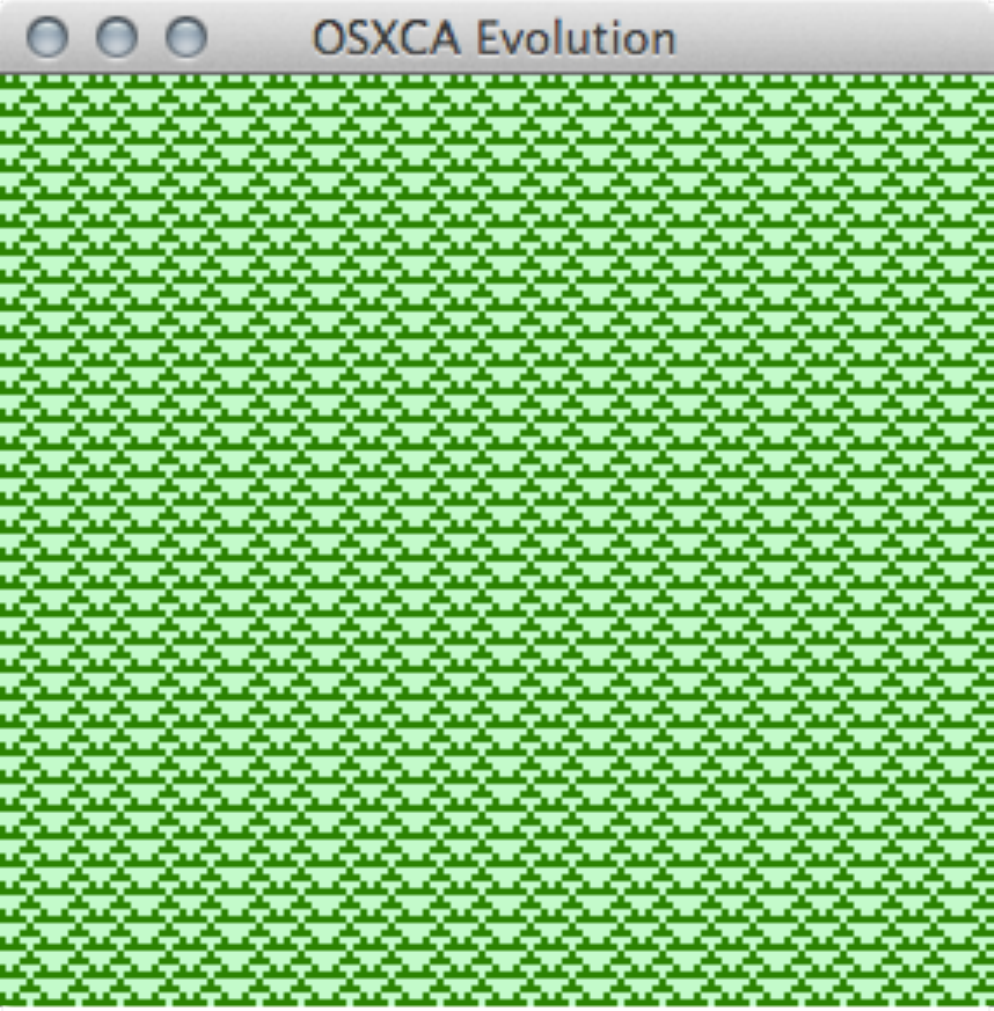}}} \hspace{0.8cm}
\subfigure[]{\scalebox{0.4}{\includegraphics{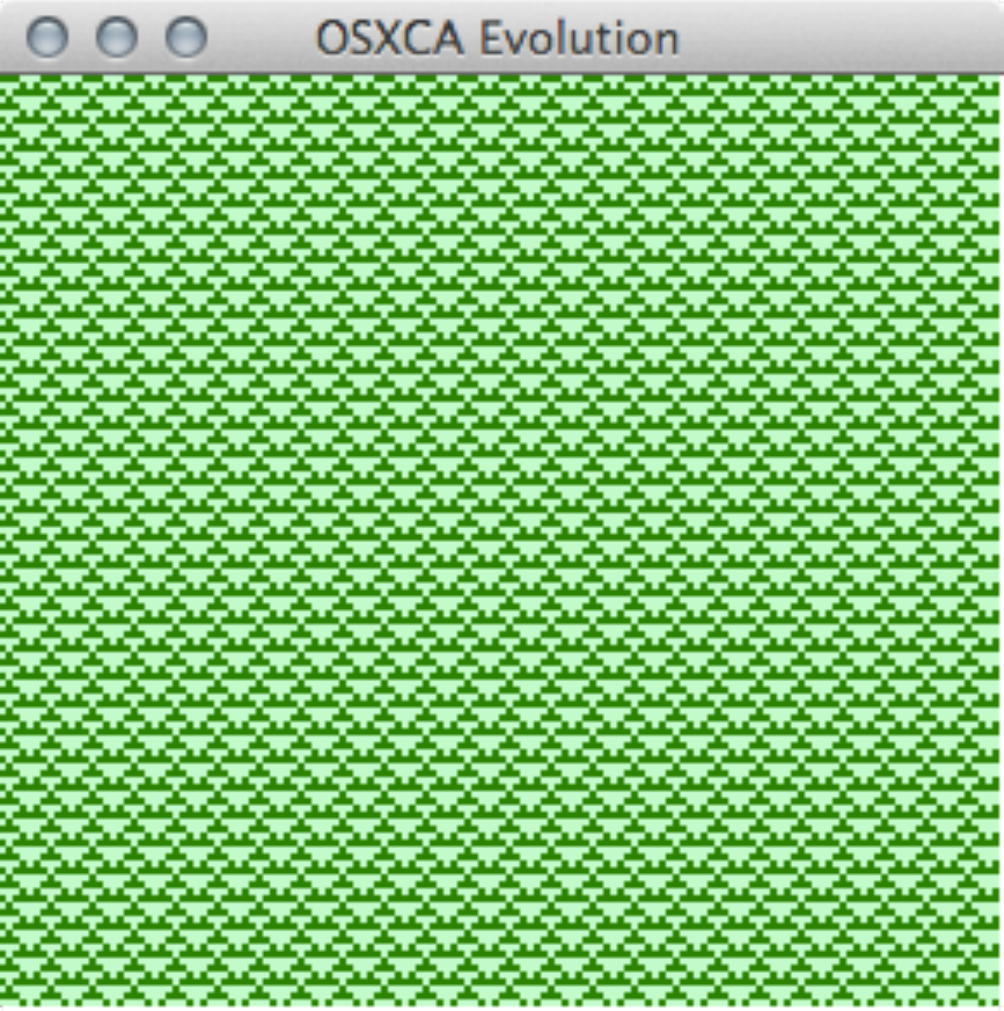}}} \hspace{0.8cm}
\subfigure[]{\scalebox{0.41}{\includegraphics{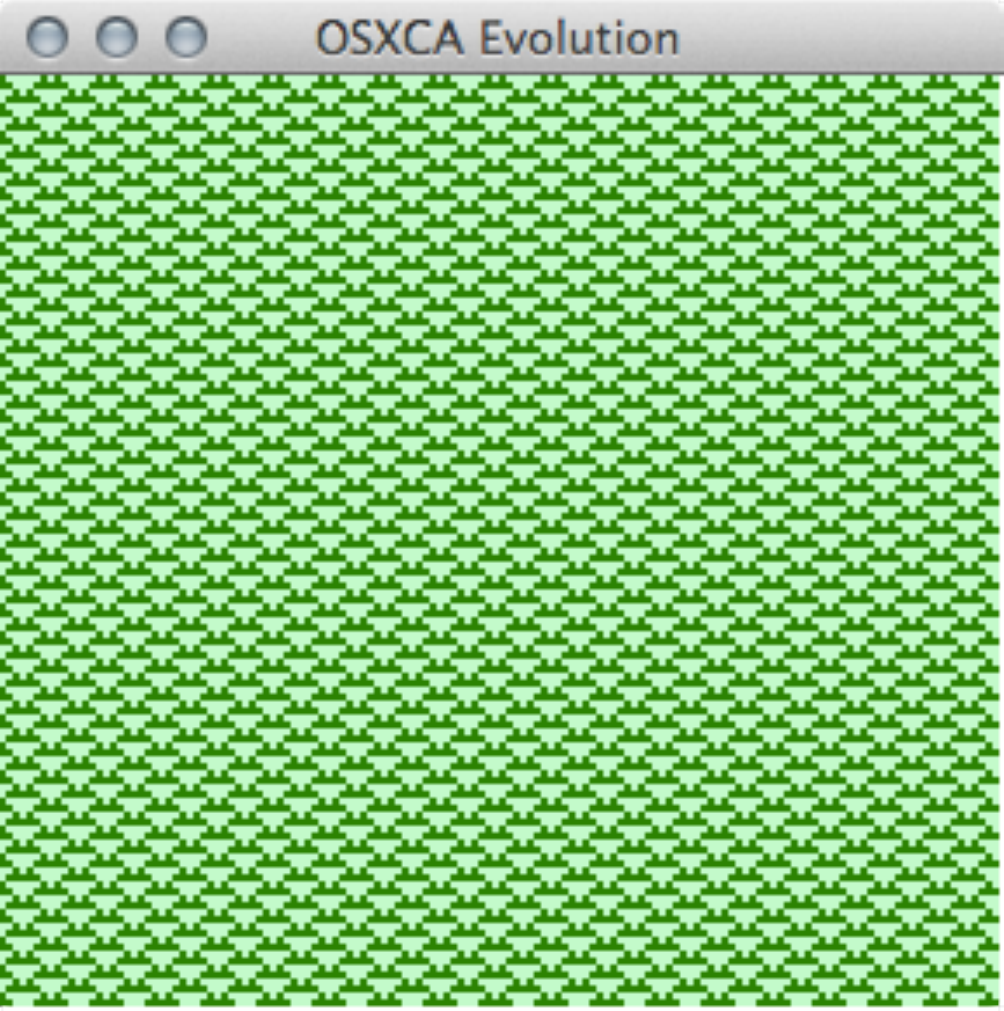}}} \hspace{0.8cm}
\subfigure[]{\scalebox{0.42}{\includegraphics{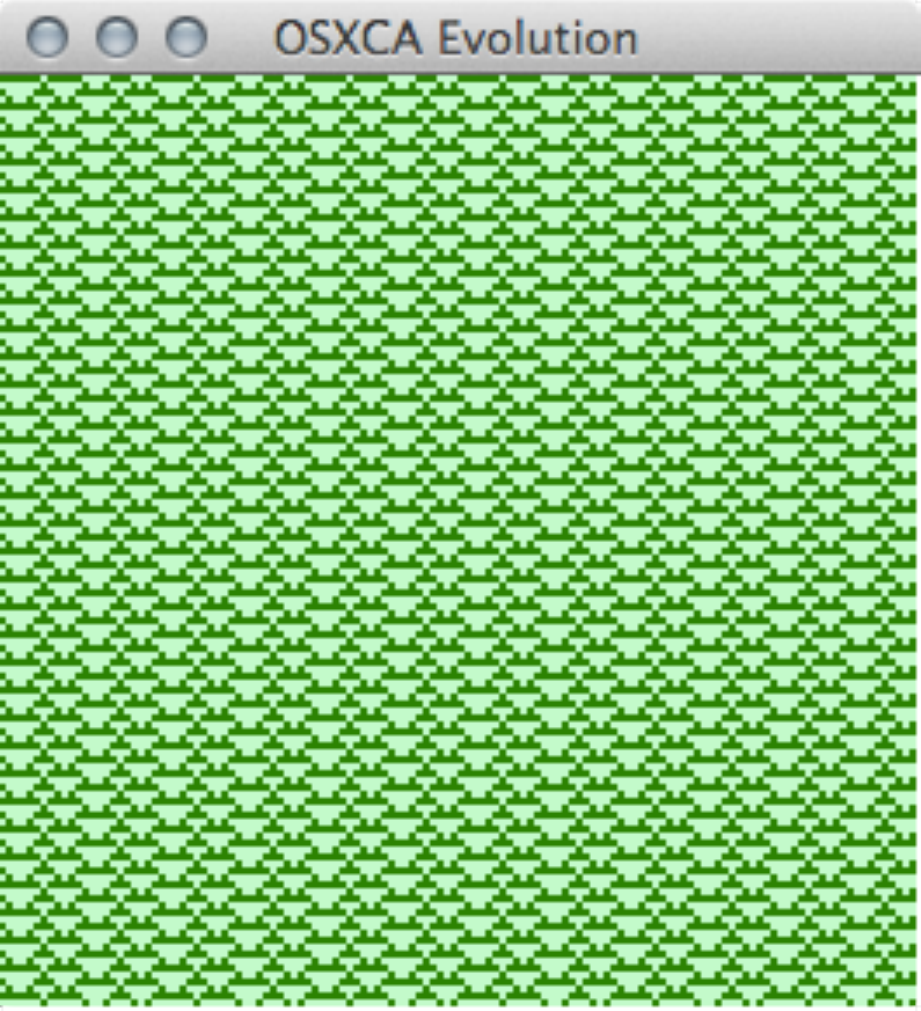}}}
\end{center}
\caption{Evolutions encoded from cycles in the de Bruijn diagram shown in  Figure~\ref{deBruijn-2}.}
\label{deBruijnCycles}
\end{figure}

\begin{itemize}
\item String $1010001001000$ encodes $g_o$-$g_e$ where both gliders are in phase three (f$_3$), and have periodic background in phase one ($e_1$) (Figure~\ref{deBruijnCycles}(a)).
\item String $111000$ encodes $g_o$ glider in phase four (f$_4$) with a periodic background in phase two ($e_2$) (Figure~\ref{deBruijnCycles}(b)).
\item String $10000010$ encodes trains of two $g_o$ gliders covering the whole evolution space. To reach this configurations it is necessary to use two different phases, f$_1$ and f$_3$ of $g_o$ glider (Figure~\ref{deBruijnCycles}(c)).
\item String $1111110111110000111000$ produces a sophisticated pattern with singular and compound gliders, and periodic background 
$g_e$-$g_o$-$g_e$-$e_2$-$g_o$ (Figure~\ref{deBruijnCycles}(d)).
\end{itemize}

\begin{figure}
\centerline{\includegraphics[width=3.3in]{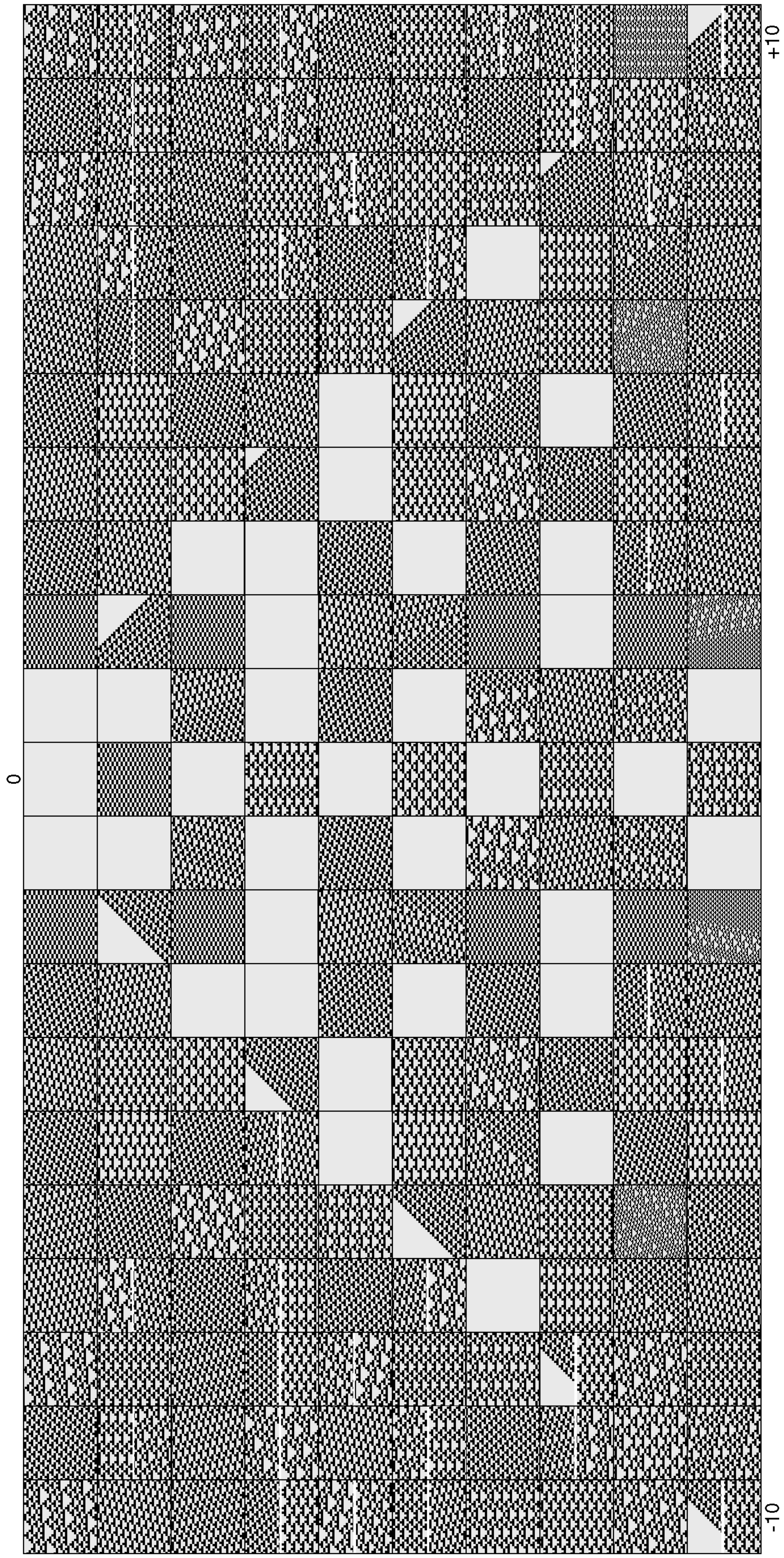}}
\caption{Periodic patterns in rule 54 calculated with extended de Bruijn diagrams for 10 generations. Each square $(x,y)$ (small snapshot evolution) displays its respective pattern.}
\label{deBruijnSmall}
\end{figure}

Thus, we can calculate systematically all periodic patterns for each $(x,y)$-position in the de Bruijn diagrams. Figure~\ref{deBruijnSmall} shows the full evolutions to 10 generations. Indeed, symmetries are preserved during its evolutions with displacements, and some positions are dominated by the stable state. Of course, we can find the periodic background and basic gliders in several positions where they match with this period and other interesting periodic patterns that emerge in rule 54.

\begin{table}
\footnotesize
\centering
\begin{tabular}{|l|l|l|}
  \hline
  $e$ & $\overrightarrow{w}$ & $\overleftarrow{w}$ \\
  \hline
  $e_1$ = 1000 & $\overrightarrow{w}($f$_1)$ = $e_1$-10-$e_2$ & $\overleftarrow{w}($f$_1)$ = $e_1$-$e_2$ \\
  $e_2$ = 1110 & $\overrightarrow{w}($f$_2)$ = $e_2$-00-$e_1$ & $\overleftarrow{w}($f$_2)$ = $e_2$-$e_1$ \\
   & $2\overrightarrow{w}($f$_1)$ = $e_1$-10111000-$e_1$ & $2\overleftarrow{w}($f$_1)$ = $e_1$-11101000-$e_1$ \\
   & & $2\overleftarrow{w}($f$_2)$ = $e_2$-11101000-$e_2$ \\
  \hline \hline
  & $g_{o}$ & $g_{e}$ \\
  \hline
  & $g_o$(A,f$_1$) = $e_1$-100000-$e_1$ & $g_e$(A,f$_1$) = $e_1$-1000000-$e_1$ \\
  & $g_o$(A,f$_2$) = $e_2$-111110-$e_2$ & $g_e$(A,f$_2$) = $e_2$-000-$e_2$ \\
  & $g_o$(B,f$_1$) = $e_1$-10-$e_1$ & $g_e$(B,f$_1$) = $e_1$-100-$e_1$ \\
  & $g_o$(B,f$_2$) = $e_2$-00-$e_2$ & $g_e$(B,f$_2$) = $e_2$-1111110-$e_2$ \\
  \hline \hline
  & gun & \\
  \hline
  & gun(A,f$_1$) = $e_1$-1111111100-$e_1$ & \\
  & gun(A,f$_2$) = $e_2$-1000000001-$e_1$ & \\
  & gun(B,f$_1$) = $e_1$-11100000010010-$e_2$ & \\ 
  & gun(C,f$_1$) = $e_1$-10001000011100-$e_2$ & \\ 
  & gun(C,f$_2$) = $e_2$-010001-$e_1$ & \\

& gun(D,f$_1$) = $e_1$-1111010010-$e_2$ & \\
& gun(D,f$_2$) = $e_2$-1000011111-$e_1$ & \\ 
& gun(E,f$_2$) = $e_2$-11100100000010-$e_2$ & \\ 
& gun(A2,f$_1$) = $e_1$-10001111000011-$e_1$ & \\ 
& gun(A2,f$_2$) = $e_2$-10000100-$e_1$ & \\ 

& gun(B2,f$_1$) = $e_1$-111001111110-$e_2$ & \\
& gun(C2,f$_1$) = $e_1$-11000000-$e_1$ & \\
& gun(C2,f$_2$) = $e_2$-10010000-$e_2$ & \\
& gun(D2,f$_1$) = $e_1$-11111100-$e_1$ & \\
& gun(D2,f$_2$) = $e_2$-100000011110-$e_2$ & \\

& gun(E2,f$_1$) = $e_1$-111000010000-$e_1$ & \\
& gun(A3,f$_1$) = $e_1$-10011100-$e_2$ & \\
& gun(A3,f$_2$) = $e_2$-11110001-$e_1$ & \\
& gun(B3,f$_1$) = $e_1$-100001010010-$e_2$ & \\
& gun(B3,f$_2$) =  $e_2$-01111111-$e_1$ & \\

& gun(C3,f$_1$) = $e_1$-110000000010-$e_2$ & \\
& gun(C3,f$_2$) = $e_2$-100100000011-$e_1$ & \\
& gun(D3,f$_1$) = $e_1$-1111110000-$e_1$ & \\
& gun(D3,f$_2$) = $e_2$-1000000100-$e_2$ & \\
& gun(E3,f$_1$) = $e_1$-1110000111-$e_1$ & \\

& gun(E3,f$_2$) = $e_2$-10001001000010-$e_2$ & \\
& gun(A4,f$_2$) = $e_2$-1111110011-$e_1$ & \\
& gun(B4,f$_1$) = $e_1$-10000001-$e_1$ & \\
& gun(B4,f$_2$) = $e_2$-00010010-$e_2$ & \\
& gun(C4,f$_1$) = $e_1$-10011111-$e_1$ & \\

& gun(C4,f$_2$) = $e_2$-111100000010-$e_2$ & \\
& gun(D4,f$_1$) = $e_1$-01000011-$e_1$ & \\
& gun(D4,f$_2$) = $e_2$-011100-$e_1$ & \\
& gun(E4,f$_1$) = $e_1$-110001-$e_2$ & \\
& gun(E4,f$_2$) = $e_2$-1001010000-$e_1$ & \\
\hline
\end{tabular}
\caption{Set of regular expressions for gliders in rule 54.}
\label{regularExpressionsR54}
\end{table}

The complete description of regular expressions representing gliders in rule 54 is provided in Table~\ref{regularExpressionsR54}. This set of regular expressions is implemented in OSXLCAU21 system \cite{kn:MarOSX}.

\subsection{The Scalar Diagram in Rule 54}

The scalar subset diagram is derived from the de Bruijn diagram. The scalar subset diagram  represents an abstract machine to verify what sequences belong to the language produced by rule 54. Also the diagram can calculate {\it Garden of Eden} configurations and other properties, as was demonstrated by McIntosh in \cite{kn:Mc91, kn:Mc09}. A Garden of Eden configuration is a configuration that cannot be achieved from any other configuration in evolution of a CA. This is a configuration without ancestors.

The subset diagram has $2^{k^{2r}}$ vertices. If all the configurations of certain length have ancestors, then all extended (with additional cells added on both ends) configurations must have ancestors. Otherwise, they describe configurations in the Garden of Eden and represent paths going from the maximum set to the minimum one.

Nodes are grouped into subsets. A note should be composed of the subsets that can be arrived at through systematic departures from all the nodes in any given subset. The result is a new graph, with subsets for nodes and links summarizing all the places that can be traveled to from all the different combinations of starting points. Sometimes, but far from always, the possible destinations narrow down as progress is made; in any event, all the possibilities have been cataloged.

Let us define the subset diagram following~\cite{kn:Mc91, kn:Mc09}. Let $a$ and $b$ be vertices, $S$ a subset and $|S|$ the cardinality of $S$. Then the subset diagram is defined as follows:

\begin{equation}
	\sum_{i} (S) = \left\{\begin{array}{lll}
				\phi & S=\phi \\ 
       		                  \{b\ |\ \mbox{edge}_{i}\ (a,b)\} & S=\{a\} \\
				\bigcup_{a \in S} \Sigma_{i}(a) & |S|>1.
		         \end{array} \right. 
\end{equation}

Three important properties are given here:

\begin{enumerate}
\item If there is a path from the maximum subset to the minimum one, then there exists a similar path starting from some smaller subset to the empty one. On the other hand, if all the unitary classes do not have edges going to the empty set, then there are no configurations in the Garden of Eden. 
\item  Given an origin and a destination, there is always a subset containing the accessible destination and another subset containing the origin; also, the destination can have additional vertices.
\item The subset diagram is not connected, and it is interesting to know the accessible greatest subset as well as the smallest one from a given subset.
\end{enumerate}

The important convention in constructing the diagram is that if it seems  there should be a link towards a certain node and if there is no such link, the link must be drawn to the empty set instead. This convention assures every label of having a representation at every node in the subset diagram.

Vertices of the subset diagram are formed by the combination of each subset formed from the states of the de Bruijn diagram (a power set). Below we discuss de Bruijn diagram -- expressing the local function $\varphi$ --  symbolized in two matrices \cite{kn:Mc09}.

Symbolic de Bruijn matrices $D_{k,s}$ or $D_{s}$ are characterized by $k$ states and $s$ number of states in the partial neighborhood. Thus for rule 54 we have the following symbolic matrices:

\[ 
	\left. \begin{array}{lll}
    	   	 D_{2,2} = \left[ \begin{array}{cccc}
		    		0 & 1 & . & . \\
	            		. & . & 1 & 1 \\
   	            		0 & 1 & . & . \\
	            		. & . & 1 & 0 \\
		   	    \end{array}
	   	\right]
	&
		             = \left[ \begin{array}{cccc}
		    		0 & . & . & . \\
	            		. & . & . & . \\
   	            		0 & . & . & . \\
	            		. & . & . & 0 \\
		   	    \end{array}
	   	\right]
	&
    	   	 + \left[ \begin{array}{cccc}
		    		. & 1 & . & . \\
	            		. & . & 1 & 1 \\
   	            		. & 1 & . & . \\
	            		. & . & 1 & . \\
		   	    \end{array}
	   	\right].
	\end{array}
	\right. 
\]

Therefore, for any ECA order $(2,1)$ we have four sequences of states in the Bruijn diagram enumerated as $\{0\}$, $\{1\}$, $\{2\}$ and $\{3\}$. All the possible subsets are $\{0,\ 1,\ 2,\ 3\}$, $\{0,\ 1,\ 2\}$, $\{0,\ 1,\ 3\}$, $\{0,\ 2,\ 3\}$, $\{1,\ 3,\ 2\}$, $\{0,\ 1\}$, $\{0,\ 2\}$, $\{0,\ 3\}$, $\{1,\ 2\}$, $\{1,\ 3\}$, $\{3,\ 2\}$, $\{3\}$, $\{2\}$, $\{1\}$, $\{0\}$ and $\{\}$. In these subsets, four unitary classes can be distinguished; the incorporation of the empty set guarantees that all subsets have at least one image, although this one does not exist in the original diagram.

\begin{table}
\centering
\begin{tabular}{|c|c|c|c|}
\hline
Subset & Node & Link with 0 & Link with 1 \\
\hline \hline
 	0,1,2,3 & 15 & 9 & 14  \\
 	1,2,3 & 14 & 9 & 14 \\
 	0,2,3 & 13 & 9 & 6 \\
 	0,1,3 & 11 & 9 & 6 \\
 	0,1,2 & 7 & 1 & 14 \\
  	2,3 & 12 & 9 & 6 \\
 	1,3 & 10 & 8 & 12 \\
 	1,2 & 6 & 1 & 14 \\
	0,3 & 9 & 9 & 6 \\
 	0,2 & 5 & 1 & 2 \\
 	0,1 & 3 & 1 & 14 \\
 	3 & 8 & 8 & 4 \\
 	2 & 4 & 1 & 2 \\
 	1 & 2 & 0 & 12 \\
 	0 & 1 & 1 & 2 \\
 	$\phi$ & 0 & 0 & 0 \\
\hline
\end{tabular}
\caption{Relation between states of the subset diagram in rule 54.}
\label{relationsubsetR54}
\end{table}

\begin{figure}
\centerline{\includegraphics[width=2.3in]{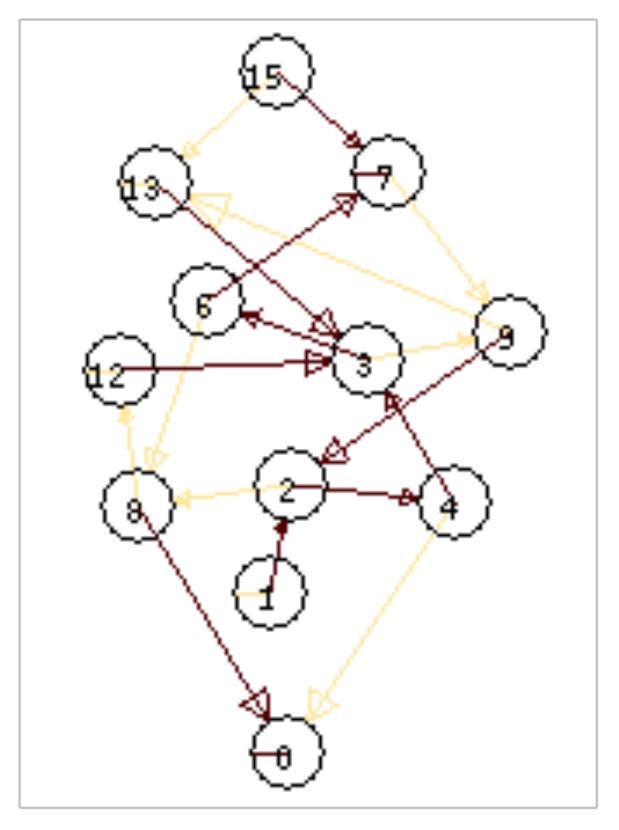}}
\caption{The scalar subset diagram of rule 54.}
\label{subSetR54color}
\end{figure}

In order to determine the type of union between the subsets, the state in which each sequence evolves must be reviewed to know toward which states (subset that form it) this subset can be connected; this way the relation for rule 54 is constructed in Table~\ref{relationsubsetR54}. The corresponding scalar subset diagram for rule 54 is shown in Figure~\ref{subSetR54color}.

Each connection is defined from its relation between subsets (see Table~\ref{relationsubsetR54}). We must distinguish four levels of subsets. Also, we should observe that a residual of the de Bruijn diagram can be founded in the subset diagram. This is because a unit class is precisely the nodes of original diagram.

At first glance, we can see that some relations is more frequent than others. There are nodes without any inputs, or nodes with most types of connections including self-loops. However, more interesting are cycles of different lengths. They are important to recognize words or sequences that a CA could recognize, as a general machine for this language.

A small subset diagram may be deduced from its original diagram. This diagram shall include only vertices with cycles, the universal and empty set, and the subset of cardinality one, yielding a new diagram that will be more practical for our proposes. The reduction gives a yet smaller diagram shown in the Figure~\ref{subSetR54color}.

Once the subset diagram has been formed, if a path leads from the universal set to the empty set, that is conclusive evidence that such a path exists nowhere in the original diagram.

\subsubsection{Garden of Eden Configurations in Rule 54}

We know that the local function $\varphi$ of rule 54 has an injective correspondence exploring its subset diagram. With this correspondence, we can find paths in the subset diagram representing Garden of Eden configurations. In this manner, we can obtain two minimal configurations that calculate Garden of Eden configurations for rule 54 represented by the strings $101010$ and $01010$. Of course, concatenations and compositions of these strings will produce a more extended Garden of Eden configuration.

\subsubsection{An Abstract Machine for Rule 54}

A practical application of the subset diagram is that it can recognize any valid string in rule 54. Another way to verify if a string derived from the de Bruijn diagram, cycle diagram, or tiles representation is to evaluate such string in the subset diagram, in the same way as regular language is recognised in classic automata theory \cite{kn:HU79, kn:Mins67}.

In order to verify this property, it is necessary to take a sequence from the set of regular expressions $\Psi_{R54}$ and check for a route match into the subset diagram. Otherwise, if such a string does not follow any route then it does not belong to $L_{R54}$.

\section{Cycles Diagrams}

\begin{figure}[th]
\centerline{\includegraphics[width=4.2in]{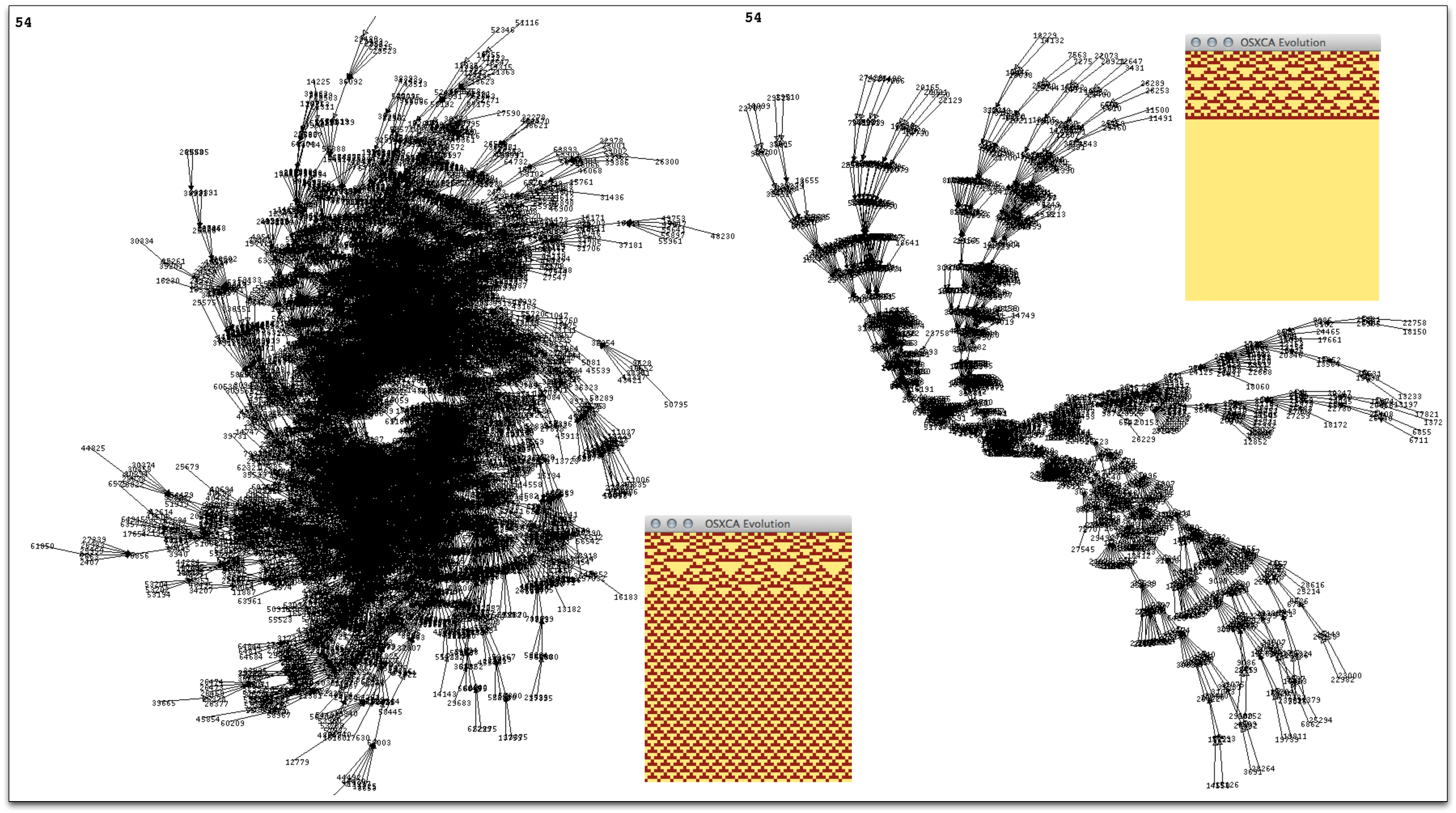}}
\caption{Cycle diagrams calculating periodic background from their attractors with $l=16$ (left) and $l=15$ (right).}
\label{cycleDiagram}
\end{figure}

Another way to get periodic structures in rule 54 is to calculate cycle diagrams (or attractors), similar to what Wuensche~\cite{kn:WL92} did by deriving an ECA classification based in basins of attraction properties.

In this section, we explore some cases with particular evolutions or attractors.

Figure~\ref{cycleDiagram} (left) determines a cycle diagram for a configuration with 16 cells. This attractor has a root cycle of four states with a total of 6432 vertices. If you choose a leaf (vertex 50795), then it is the periodic configuration that will evolve during 32 generations to reach the attractor, which is precisely the periodic background configurations. Figure~\ref{cycleDiagram} (right) determines a cycle diagram for a configuration with 15 cells, it has an attractor with just one state, the stable state, that can be reached after 21 generations starting with the configuration vertex 11491, this attractor has 1583 vertices.

\begin{figure}
\centerline{\includegraphics[width=4.1in]{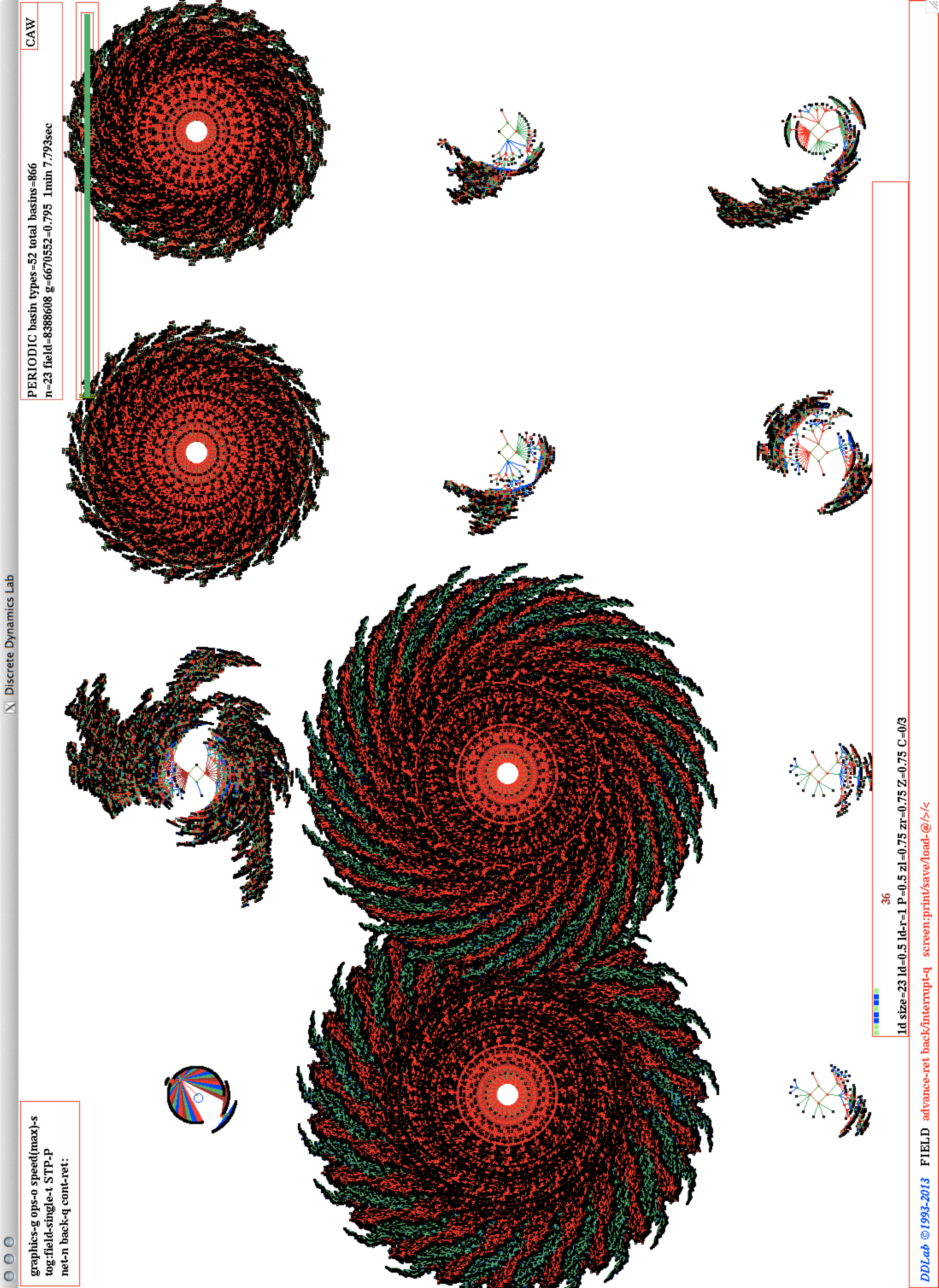}}
\caption{Basin of attractors in rule 54 for rings with 23 cells.}
\label{attractorsS23R54}
\end{figure}

Figure~\ref{attractorsS23R54} displays a basin of attraction for configurations with 23 cells. We show this attractor to demonstrate the complex behavior of rule 54. It is determined by asymmetric long-transient attractors. They imply the existence of gliders and nontrivial behavior.

\begin{figure}
\centerline{\includegraphics[width=4.2in]{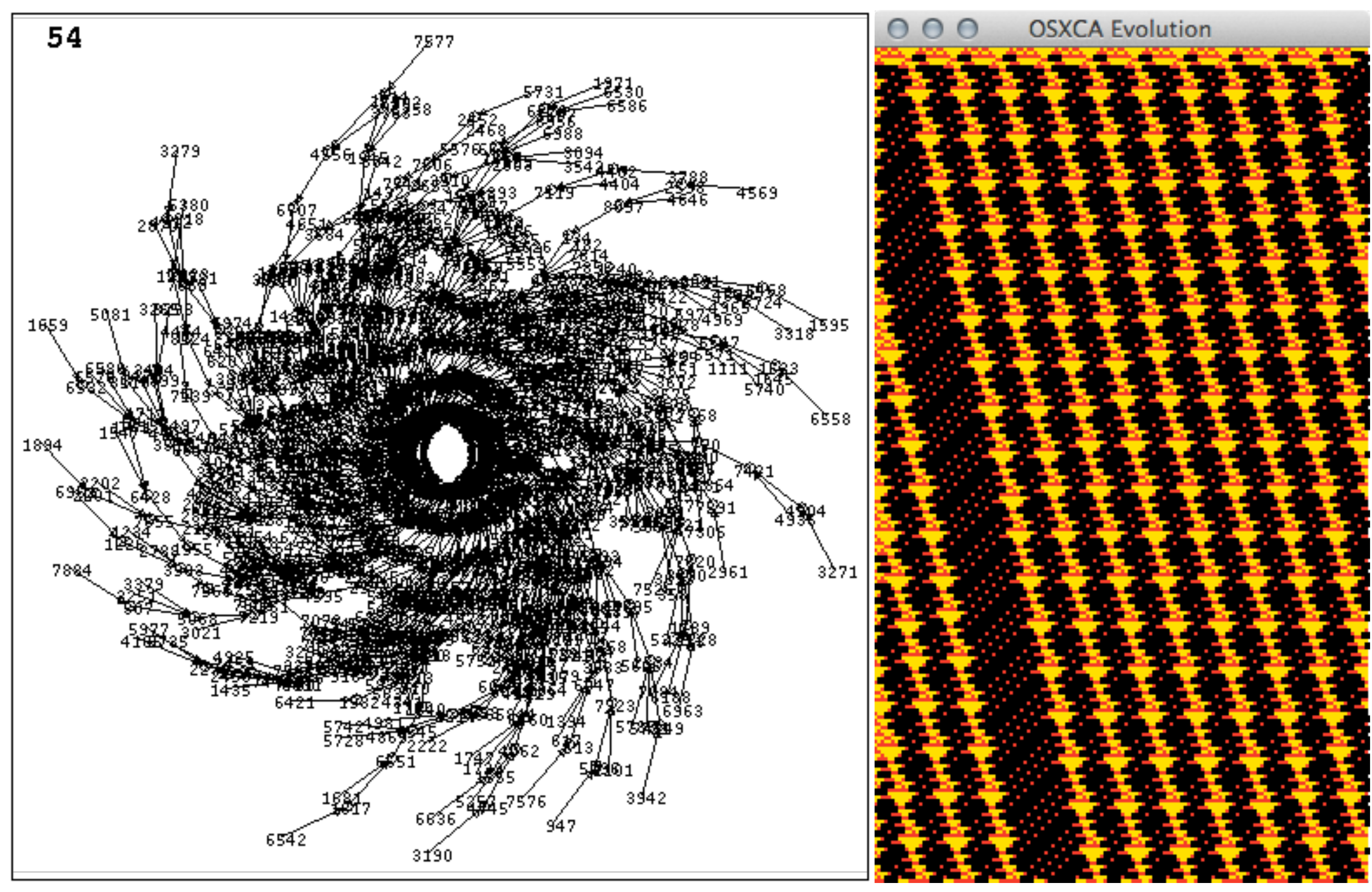}}
\caption{A cycle diagram 13 cells calculating a meta-glider or agar configuration in rule 54.}
\label{cycleDiagram-2}
\end{figure}

Figure~\ref{cycleDiagram-2} shows a ``meta-glider'', meshes, or agar configuration (an agar configuration comes from Conway's CA {\it Game of Life} literature, for details see \cite{kn:agar}). The meta-glider in this figure is a periodic structure moving to the left (composed for one $T_8$, two $T_4$, and one $T_2$ tiles); it is preserved during a triple permanent collision of three $\overleftarrow{w}$ glider. We have selected vertex 7577, which needs 20 generations to reach the attractor that represents this meta-glider, which corresponds to 169 vertices. The full attractor is composed for 1274 vertices.

For the following cycles or attractors diagrams we can list a number of periodic strings as well. In this case, every primitive glider may be reproduced from different cycles, as Table~\ref{cyclesRelationsR54} shows. The Length column indicates the attractor period, the Cycle column indicates the number of components selected that have the same cycle length, the Total Vertices column is the total number of nodes for each attractor (including branches and leaves), and the Structures column describes the number of periodic structures evolving with these strings.

\begin{table}
\small
\vspace{2pt}
\centering
\begin{tabular}{|c|c|c|l|}
\hline
Length & Cycle & Total Vertices & Structures \\[2pt]
\hline \hline
4 & 4 & 4 & $T_3$ and $T_2$ tiles \\
\hline
6 & 4 & 5 & $g_e$ glider \\
\hline
8 & 4 & 14 & $g_e$ gliders joined \\
   & 6 & 28 & $g_e$ glider with a $T_2$ \\
\hline
9 & 4 & 44 & $g_e$-$g_o$ gliders joined \\
   & 27 & 45 & $T_4$ transporting a $\overleftarrow{w}$ \\
   & & & (extensible as a $T_5$ in rule 110) \\
\hline
10 & 30 & 90 & two $T_4$ tiles joined \\
\hline
11 & 4 & 125 & $g_o$ glider with a $T_6$ tile \\
      & 11& 55 & packages of $T_4$ tiles \\
      & 99 & 231 & meta-glider ($\overrightarrow{w}$-$T_5$-$T_6$-$T_4$-$T_2$ tiles) \\
\hline
12 & 10 & 124 & periodic background (2$T_6$-2$T_3$-$T_2$ tiles) \\
      & 12 & 102 & 2$\overrightarrow{w}$ gliders \\
\hline
13 & 4 & 406 & ($g_e$-$g_o$) gliders concatenated \\
      & 169 & 1274 & meta-glider ($T_8$-2$T_4$-$T_2$ and $\overrightarrow{w}$ gliders) \\
\hline
14 & 112 & 805 & meta-glider ($T_8$-3$T_4$-$T_2$ tiles) \\
\hline
15 & 330 & 7680 & meta-glider ($T_5$-2$T_6$-$T_4$-$T_2$ tiles) \\
\hline
16 & 6 & 116 & periodic background ($T_6$-$T_2$ tiles) \\
      & 8 & 8 & $\overrightarrow{w}$ gliders \\
      & 14 & 944 & meta-glider ($\overrightarrow{w}$-$g_o$-$\overleftarrow{w}$ gliders) \\
      & 16 & 2896 & 2$\overrightarrow{w}$ gliders \\
      & 40 & 1246 & meta-glider ($T_8$-5$T_6$-2$T_2$-3$T_4$-$T_5$ tiles) \\
\hline
\end{tabular}
\caption{Cycle diagrams calculating periodic structures in rule 54.}
\label{cyclesRelationsR54}
\end{table}

\section{A Way to Encode Gliders in Rule 54}

We can encode gliders in regular expressions via the gliders' phase representations:

\begin{equation}
\#_1(\#_2,p_i),
\end{equation}

\noindent where $\#_1$ represents a glider of rule 54 of the set of gliders $\cal G$$_{R54}$, $\#_2$ represents its block of phases, and $p_i$ is a phase determined for each block of phases, where $i = \{1,2\}$. All sets of phases for gliders in rule 54 are detailed in Table~\ref{regularExpressionsR54}.

The displacement for each glider $g$ in ${\cal G}_{R54}$ is represented with the following equation:

\begin{equation}
d_{g}=2*lpm - 2*rpm.
\label{eq-desplazamiento}
\end{equation}

All periodic structures have a period length defined by the amount of margins $lpm$ and $rpm$, given its number of tiles and contact points in the structure (see Table~\ref{tablaGlidersR54}). Therefore the period of gliders is determined as

\begin{equation}
p_{g}=2*lpm + 2*rpm,
\label{eq-periodo}
\end{equation}

\noindent and the speed of gliders in rule 54 is determined as

\begin{equation}
v_{g} = \frac{2*lpm - 2*rpm}{2*lpm + 2*rpm}.
\label{eq-velocidad}
\end{equation}

Collisions between gliders have a maximum level that is determined by the number of margins $lpm$ and $rpm$, although they could not all be viable collisions. This way, a glider with $lms$ contact points and another glider with $rpm$ contact points have the next number of possible collisions:

\begin{equation}
c \leq lpm * rpm,
\label{eq-choquesmax}
\end{equation}

\noindent where $c$ represents the maximum number of possible collisions. 

Frequently, however, gliders have contact and noncontact points where the maximum level is not fulfilled. Simplifying the equation, we obtain the number of collisions between two gliders $g_{i}$ and $g_{j}$, where $i \neq j$, which is represented by the following equation:

\begin{equation}
c = |(lpm_{g_{i}} * rpm_{g_{j}}) - (rpm_{g_{j}} * lpm_{g_{i}})|.
\label{eq-choques}
\end{equation}

Therefore, following is the set of regular expressions and codification in phases for gliders in rule 54 (see Table~\ref{regularExpressionsR54}). We are able to codify easily the initial conditions to control and synchronize collisions between gliders. In the next sections we select some problems, such as, construction of gliders by collisions, unlimited growth, holes, solitons, and some simple computable devices.

\subsection{Self-Organisation by Glider Reaction}

\begin{figure}
\begin{center}
\subfigure[]{\scalebox{0.47}{\includegraphics{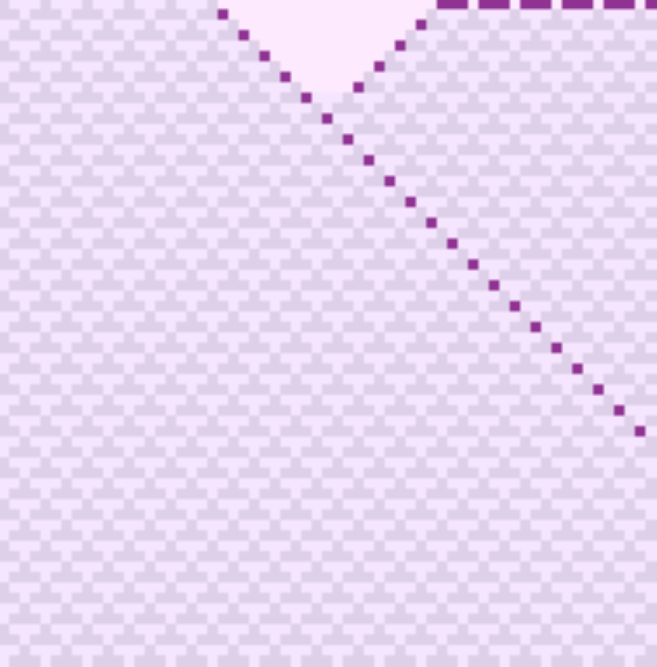}}} 
\subfigure[]{\scalebox{0.47}{\includegraphics{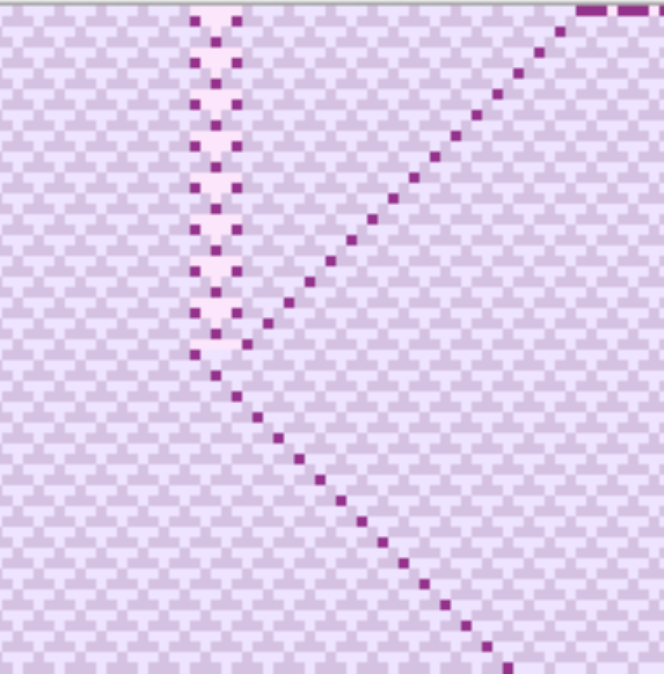}}} 
\subfigure[]{\scalebox{0.47}{\includegraphics{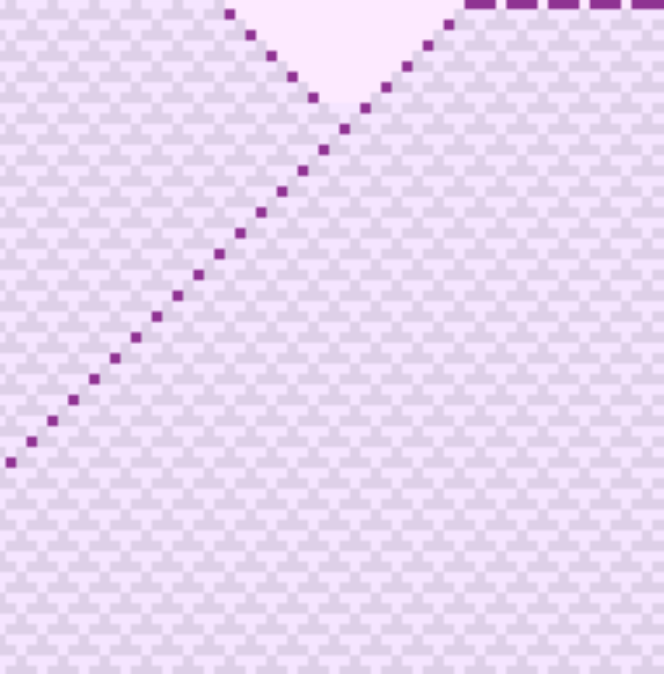}}} 
\subfigure[]{\scalebox{0.47}{\includegraphics{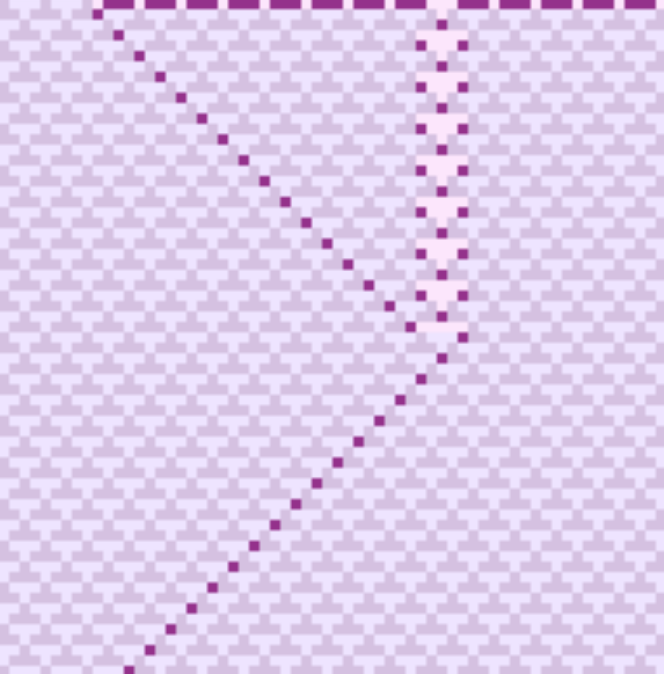}}} 
\subfigure[]{\scalebox{0.47}{\includegraphics{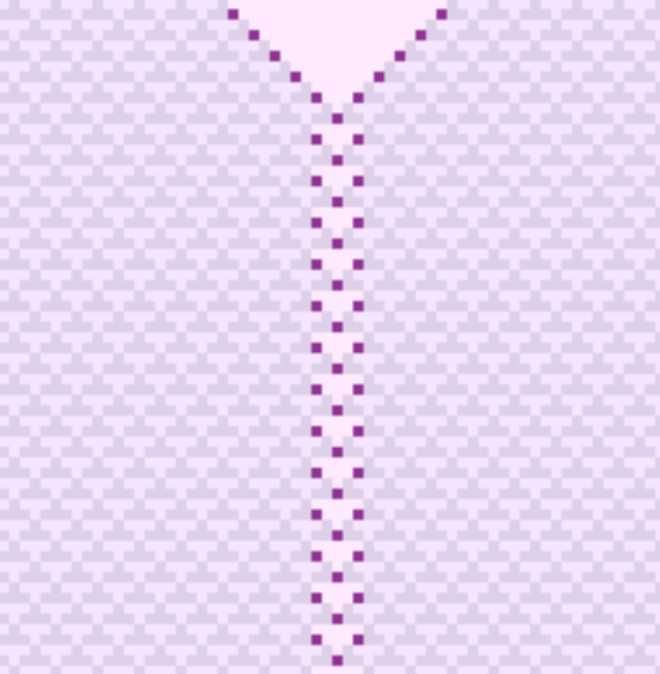}}} 
\subfigure[]{\scalebox{0.47}{\includegraphics{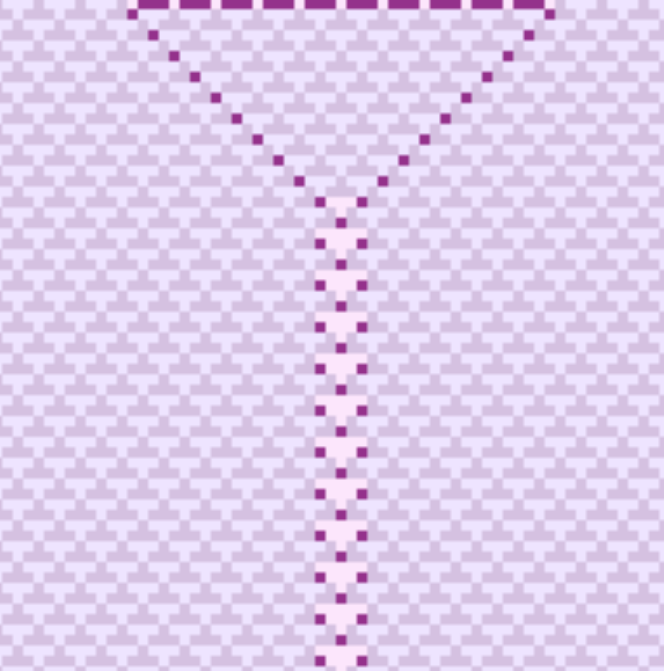}}} 
\subfigure[]{\scalebox{0.57}{\includegraphics{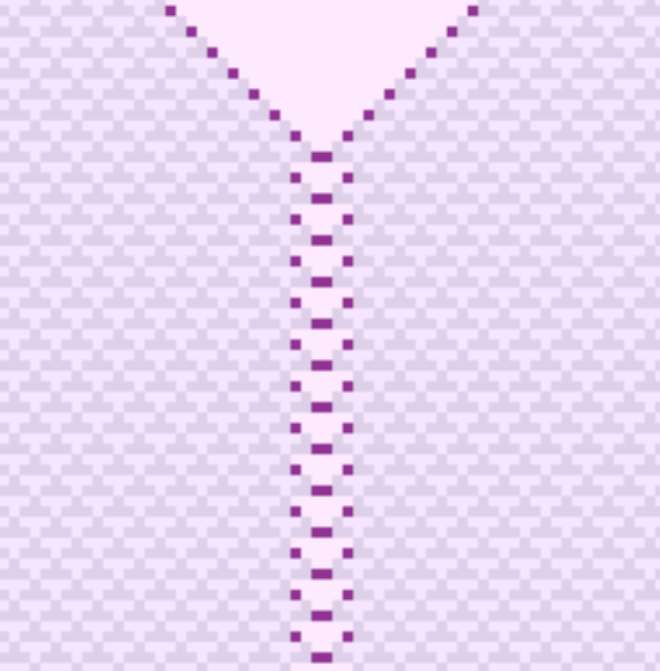}}} \hspace{0.6cm}
\subfigure[]{\scalebox{0.67}{\includegraphics{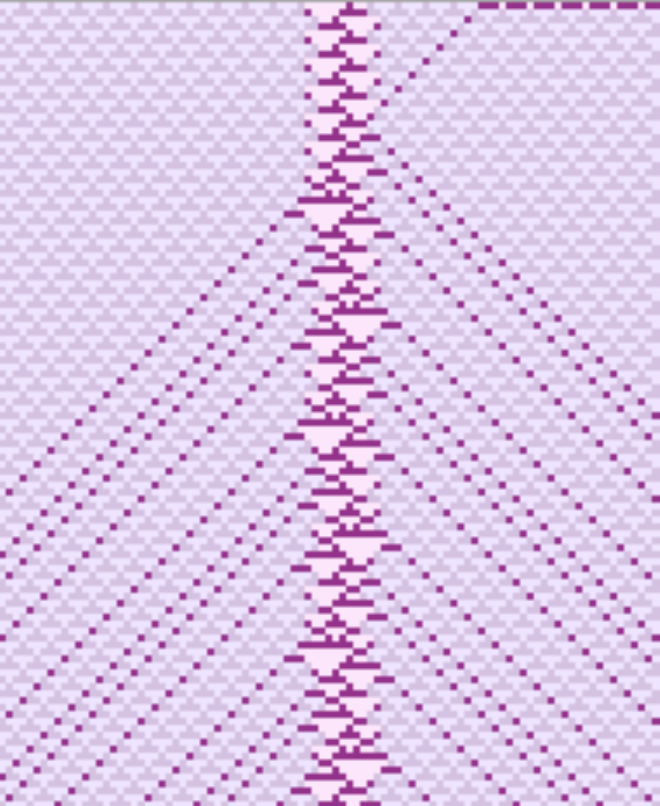}}}
\end{center}
\caption{Producing gliders by collisions in rule 54.}
\label{prodGlidersCollisions}
\end{figure}

In~\cite{kn:MAM06} we show how to construct all gliders in rule 54 from collisions between gliders. This problem is referred to as glider self-organization by collisions in complex systems \cite{kn:Kau93}. Figure~\ref{prodGlidersCollisions} displays the production of primitive gliders in rule 54, and Table~\ref{tablaGlidersCollisions} shows encoding of the collisions.

We can chooce between production by gliders or by sequences. If we want to produce a $\overrightarrow{w}$ glider, then we need collide a $g_o$ glider with a $\overleftarrow{w}$ glider and so on. We enumerate each expression to reproduce every collision presented in Figure~\ref{prodGlidersCollisions}.

\begin{enumerate}
\item $\overrightarrow{w} = ne_{1}$-$\overrightarrow{w}$-$0^{10}$-$\overleftarrow{w}$-$ne_{2}$ (Figure~\ref{prodGlidersCollisions}a).
\item $\overrightarrow{w} = ne_{1}$-($g_{o}$(A,f$_1$) $\|$ $g_{o}$(B,f$_1$))-$10e_{1}$-$\overleftarrow{w}$-$ne_{2}$ (Figure~\ref{prodGlidersCollisions}b).
\item $\overleftarrow{w} = ne_{1}$-$\overrightarrow{w}$-$0^{12}$-$\overleftarrow{w}$-$ne_{2}$ (Figure~\ref{prodGlidersCollisions}c).
\item $\overleftarrow{w} = ne_{1}$-$\overrightarrow{w}$-$8e_2$-($g_{o}$(A,f$_1$) $\|$ $g_{o}$(B,f$_1$))-$ne_{2}$ (Figure~\ref{prodGlidersCollisions}d).
\item $g_o = ne_{1}$-$\overrightarrow{w}$-$0^{10}$-$\overleftarrow{w}$-$ne_{1}$ (Figure~\ref{prodGlidersCollisions}e).
\item $g_o = ne_{1}$-$\overrightarrow{w}$-$10e_{2}$-$\overleftarrow{w}$-$ne_{1}$ (Figure~\ref{prodGlidersCollisions}f).
\item $g_e = ne_{1}$-$\overrightarrow{w}$-$0^{12}$-$\overleftarrow{w}$-$ne_{1}$ (Figure~\ref{prodGlidersCollisions}g).
\item gun = $ne_{1}$-$g_e$(A,f$_1$)-$g_e$(B,f$_1$)-4$e_{1}$-$\overleftarrow{w}$-$ne_{2}$ (Figure~\ref{prodGlidersCollisions}h).
\end{enumerate}

\noindent where $n$ is a  number of copies of the string.

Of course, different parameters will yield a glider with different intervals or a number of gliders.

\begin{table}[th]
\centering
\begin{tabular}{|c|c|c|}
\hline
           & \multicolumn{2}{c|}{Collisions} \\
\cline{2-3}
Glider & By Gliders reaction & By Sequences \\
\hline \hline
$\overrightarrow{w}$ & $g_{o}$,$\overleftarrow{w}$ & $e_{1}$*$0^{4n-2}e_{2}$* $\forall$ $n>0$ \\
\hline
$\overleftarrow{w}$ & $\overrightarrow{w}$,$g_{o}$ & $e_{1}$*$0^{4n}e_{2}$* $\forall$ $n>0$ \\
\hline
$g_{o}$ & $\overrightarrow{w}$,$\overleftarrow{w}$ & $e_{1}$*$10^{n}e_{1}$* $\forall$ $n>0$ and odd \\
\hline
$g_{e}$ & & $e_{1}$*$10^{n}e_{1}$* $\forall$ $n>0$ and even \\
\hline
glider gun & $\overrightarrow{w}$,$2g_{e}$ or $2g_{e}$,$\overleftarrow{w}$ & \\
\hline
glider gun$^{n}$ & $\overrightarrow{w}$,$g_{e}$,$2g_{e}$ or $2g_{e}$,$g_{e}$,$\overleftarrow{w}$ & \\
\hline
\end{tabular}
\caption{Collision sequence for glider production in rule 54.}
\label{tablaGlidersCollisions}
\end{table}

\subsection{Unlimited Growth}

\begin{figure}[th]
\begin{center}
\subfigure[]{\scalebox{0.37}{\includegraphics{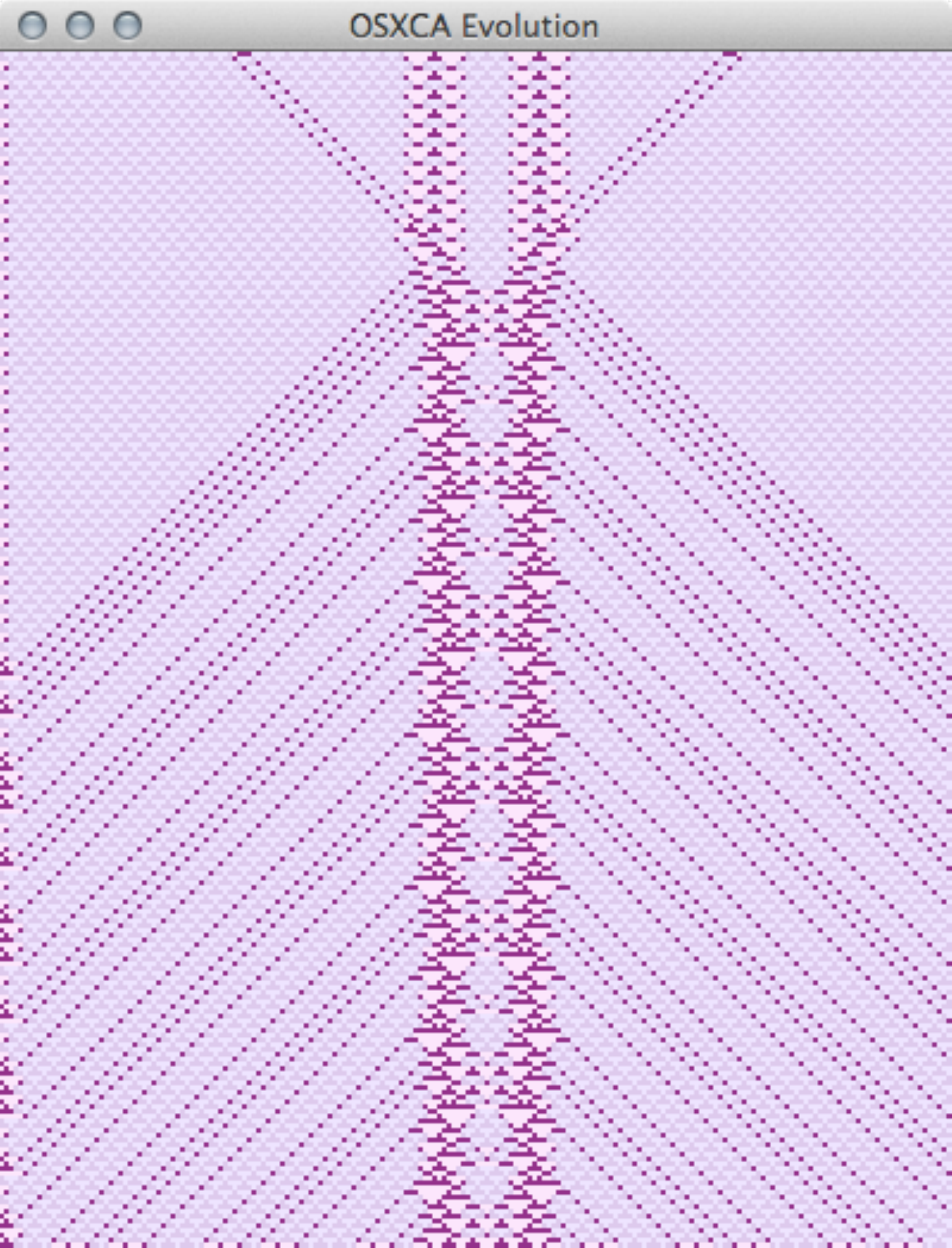}}} 
\subfigure[]{\scalebox{0.37}{\includegraphics{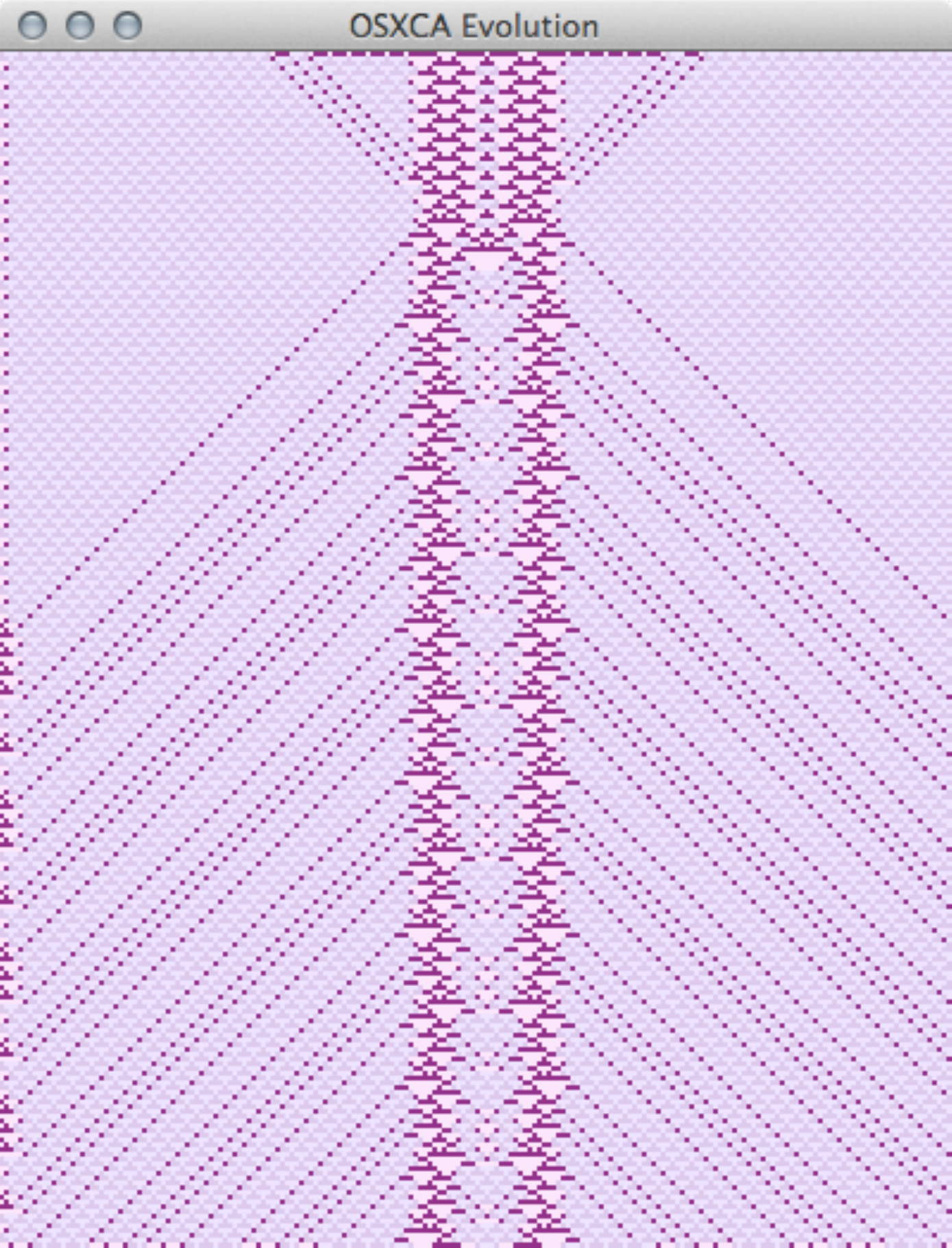}}}
\end{center}
\caption{Double glider guns in rule 54 produced from multiple collisions of (a) eight gliders, and (b) 12 gliders.}
\label{doubleGun}
\end{figure}

A famous problem established in Conway's Game of Life was discovery of a configuration that will grow permanently, into an infinite evolution space. This problem was solved by Gosper and colleagues at MIT Artificial Intelligence Lab \cite{kn:chess}.

The same problem can be established in rule 54. Of course, the construction of a glider gun or some other extension is sufficient to demonstrate unlimited growth in rule 54 (Figure~\ref{prodGlidersCollisions}h). Here we show the production of double glider guns.

\begin{enumerate}
\item Double glider gun = $ne_{1}$-$2\overrightarrow{w}$-8$e_1$-2$g_e$(A,f$_1$)-2$e_1$-2$g_e$(A,f$_1$)-8$e_1$-$2\overleftarrow{w}$ \\ -$ne_{1}$ (Figure~\ref{doubleGun}a).
\item Double glider gun = $ne_1$-$3\overrightarrow{w}$-$5e_2$-$g_e$(B,f$_2$)-$g_e$(A,f$_2$)-$g_e$(B,f$_2$)-$g_e$(B,f$_2$)- \\
$g_e$(A,f$_2$)-$g_e$(B,f$_2$)-$5e_2$-$3\overleftarrow{w}$-$ne_1$ (Figure~\ref{doubleGun}b).
\end{enumerate}

\subsection{Holes and Big Tiles}

In \cite{kn:Mc00} McIntosh determined that ECA rule 110 can be studied as a tile problem. What is a largest tile produced via collision between gliders in rule 54? Some answers are given in \cite{kn:MAM06} via studying reactions between gliders.

\begin{figure}[th]
\begin{center}
\subfigure[]{\scalebox{0.39}{\includegraphics{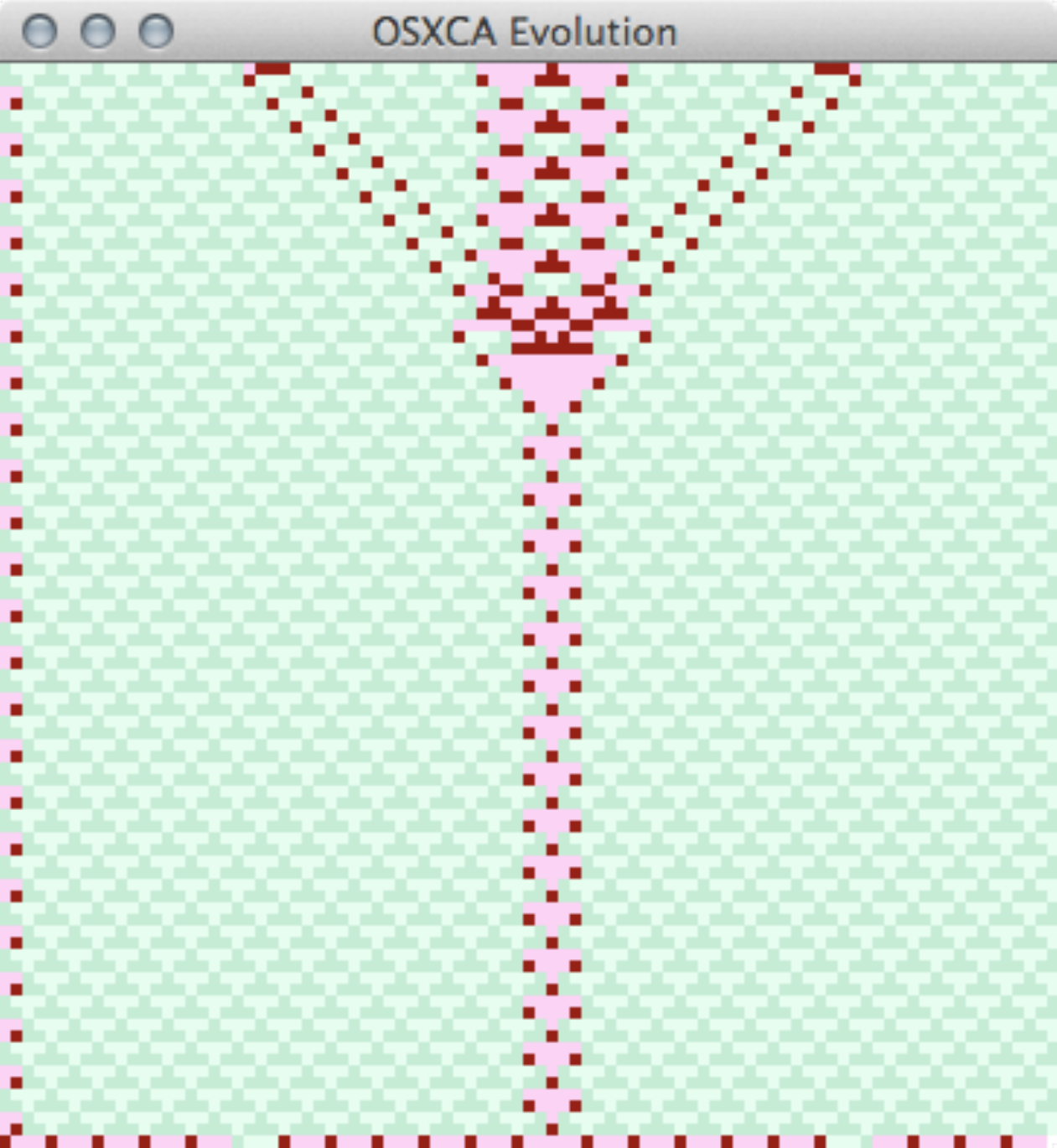}}} \hspace{0.2cm}
\subfigure[]{\scalebox{0.39}{\includegraphics{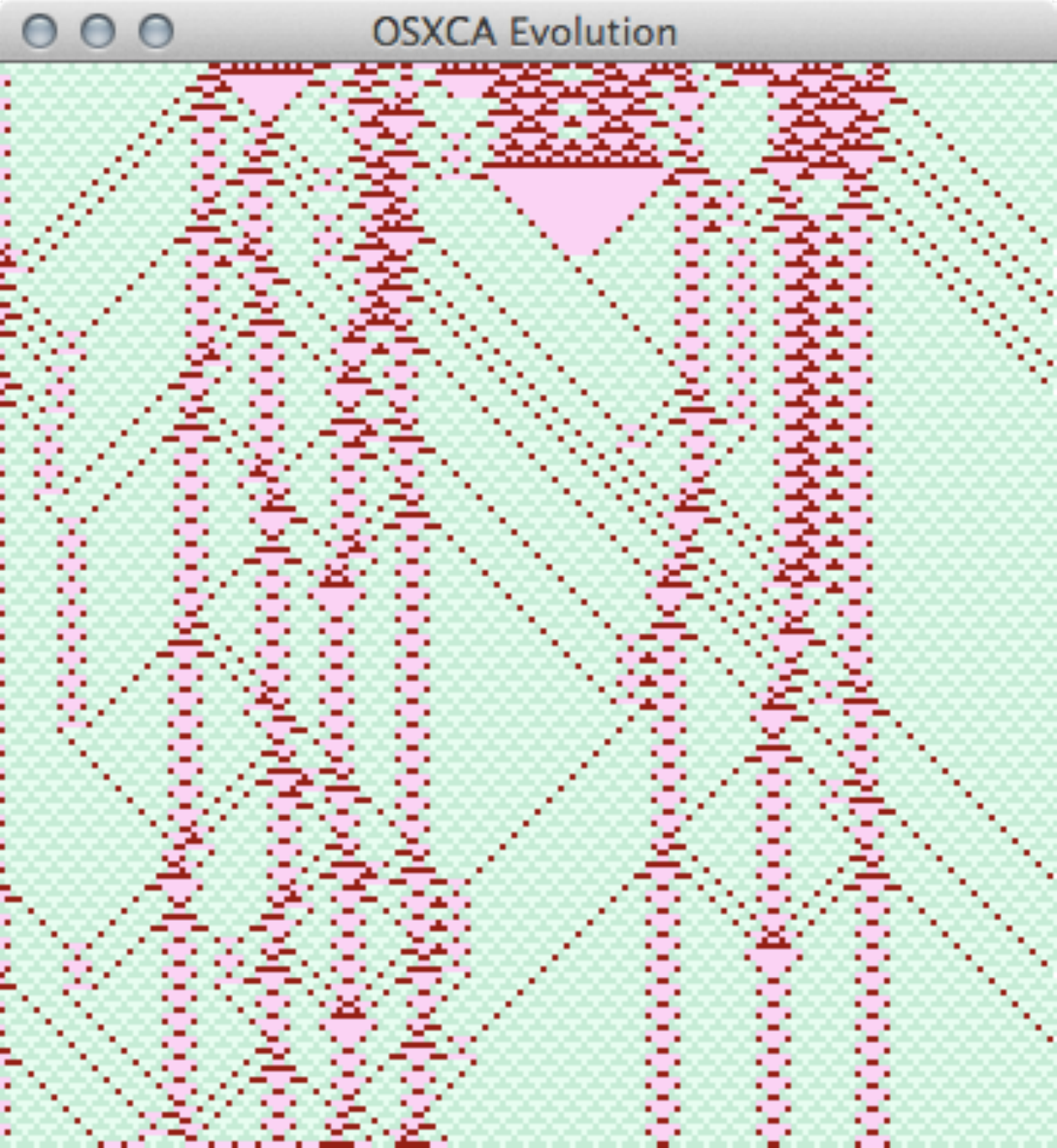}}}
\end{center}
\caption{Big tiles emerging in rule 54. (a) $T_{16}$ tile from six colliding gliders, (b) $T_{33}$ tile as a decomposition from a specific string.}
\label{bigTiles}
\end{figure}

Figure~\ref{bigTiles}(a) shows the construction of a $T_{16}$ tile by synchronising multiple collisions between $\overrightarrow{w}$, $\overleftarrow{w}$, and $g_e$ gliders. Figure~\ref{bigTiles}(b) shows  
 $T_{33}$ tile produced by a chaotic decomposition. Codes to reproduce these reactions are as follows:

\begin{enumerate}
\item $T_{16}$ = $ne_{1}$-$2\overrightarrow{w}$-4$e_{1}$-2$g_e$(A,f$_1$)-4$e_{1}$-$2\overleftarrow{w}$(A,f$_1$)-$ne_{1}$ (Figure~\ref{bigTiles}a).
\item $T_{33}$ = $ne_{1}$-110010101001010100111000001001110110000010100101001\\111011001110111001100000010101101000111010101000000101001100\\0101101000-$ne_{1}$  (Figure~\ref{bigTiles}b).
\end{enumerate}

\subsection{Memory Functions}

\begin{figure}[th]
\begin{center}
\subfigure[]{\scalebox{0.36}{\includegraphics{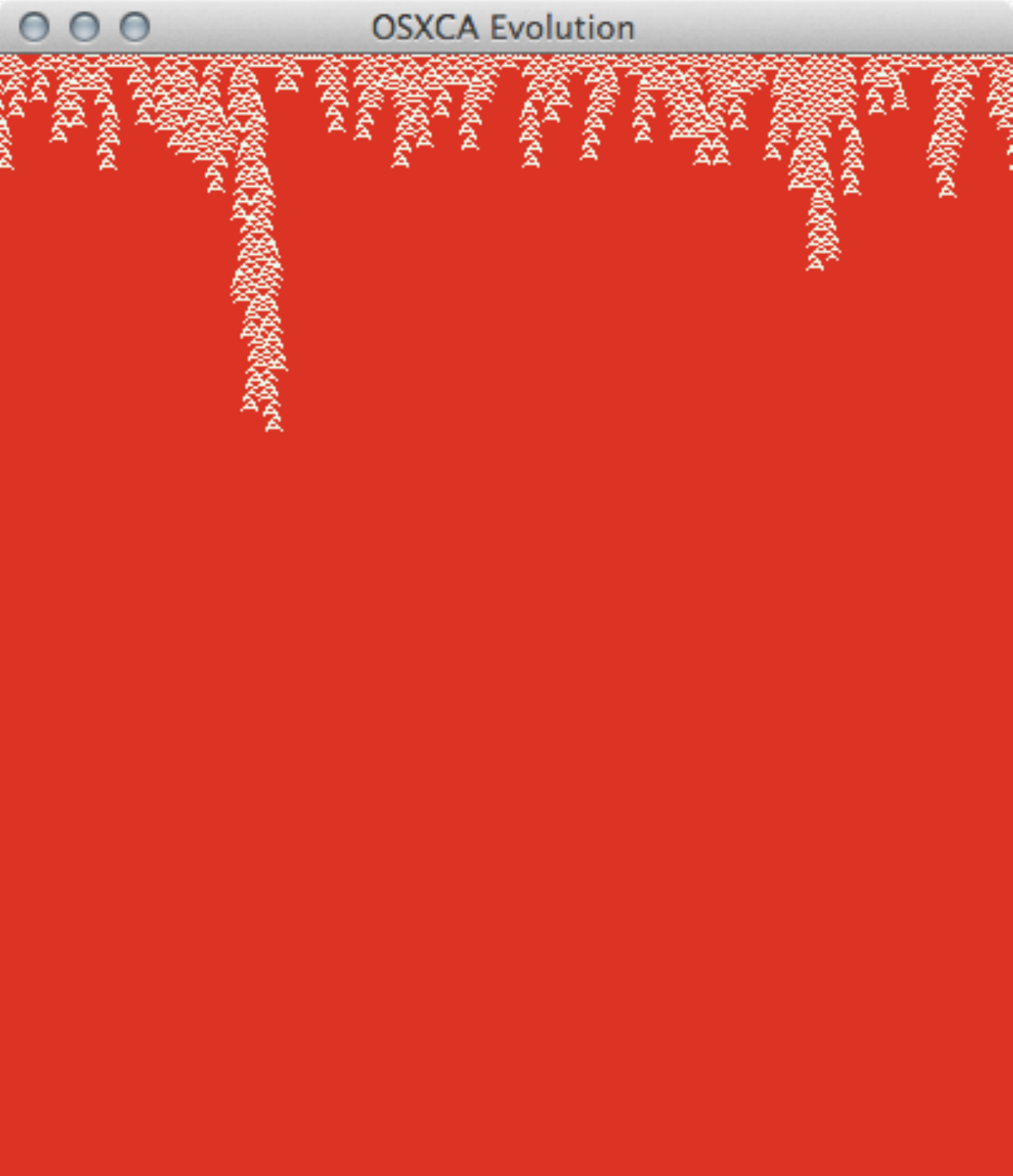}}} 
\subfigure[]{\scalebox{0.36}{\includegraphics{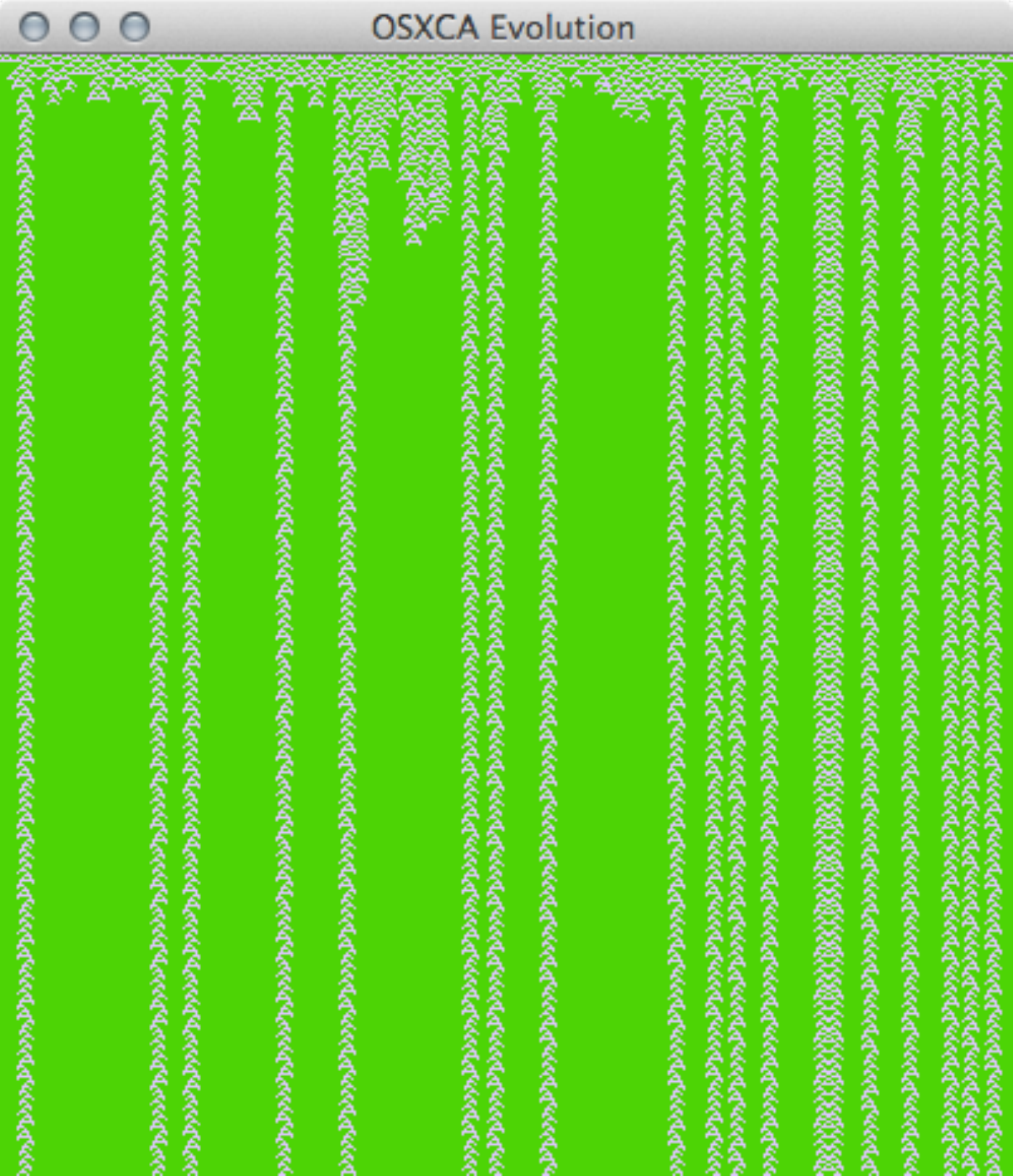}}} 
\subfigure[]{\scalebox{0.36}{\includegraphics{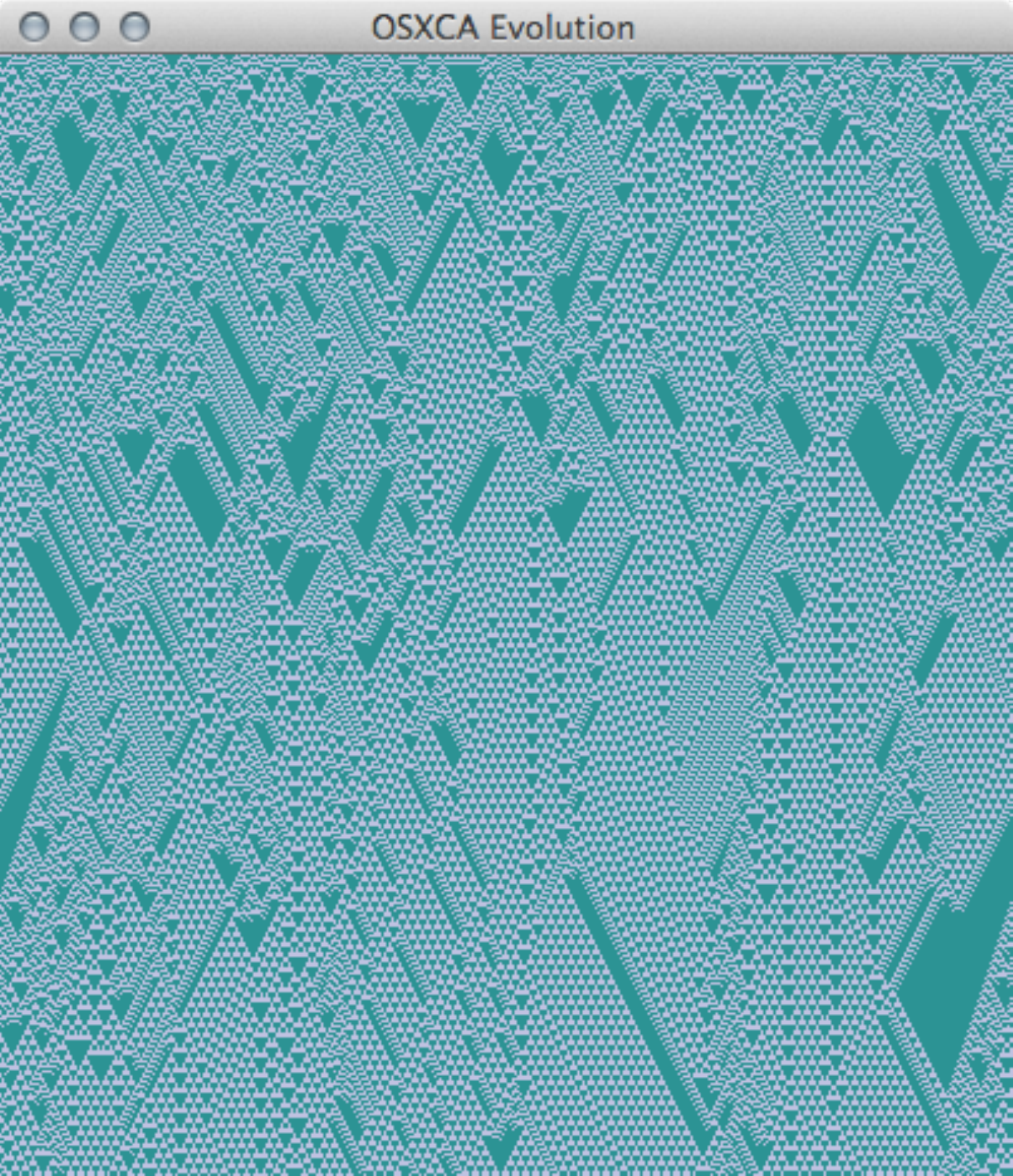}}} 
\subfigure[]{\scalebox{0.36}{\includegraphics{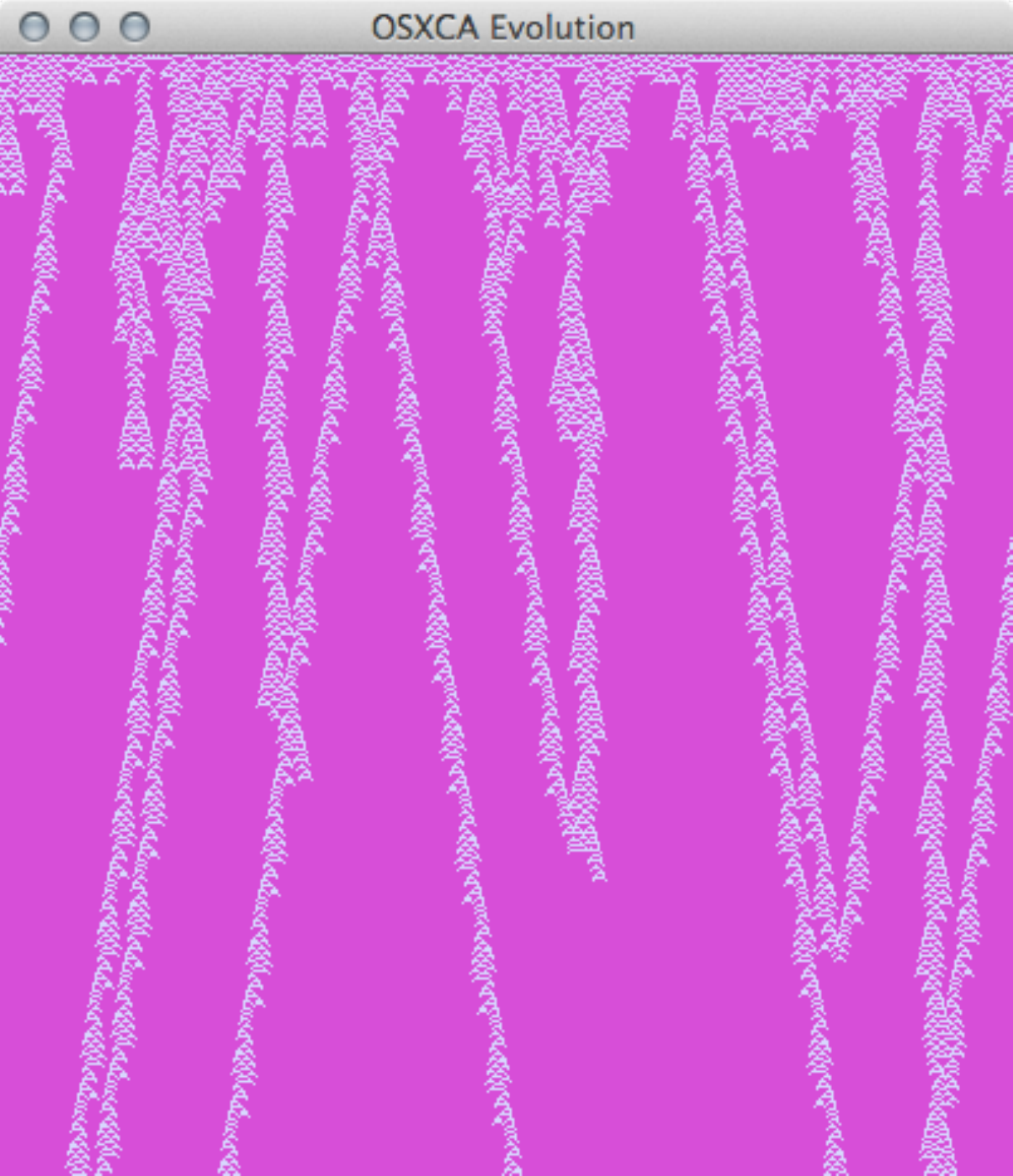}}}
\end{center}
\caption{Rule 54 affected with memory functions. (a) ECAM $\phi_{R54maj:6}$, (b) ECAM $\phi_{R54maj:10}$, (c) ECAM $\phi_{R54maj:3}$, (d) ECAM $\phi_{R54maj:8}$.}
\label{memoryR54}
\end{figure}

Rule 54 has been proved to be a `universal dynamics rule' in the ECA memory (ECAM) classification \cite{kn:MAA13}. This means that rule 54 operated with some memory functions is able to reach any Wolfram class, including the class IV, to which the memoryless rule 54 belongs~\cite{kn:Mar13}.

Figure~\ref{memoryR54} presents evolutions of rule 54 with memory. Each snapshot illustrates four different behaviors. Figure~\ref{memoryR54}(a) shows a uniform evolution with rule $\phi_{R54maj:6}$, Figure~\ref{memoryR54}(b) a periodic behavior with rule $\phi_{R54maj:10}$, Figure~\ref{memoryR54}(c) a chaotic evolution with rule $\phi_{R54maj:3}$, and Figure~\ref{memoryR54}(d) a complex behavior with rule $\phi_{R54maj:8}$. Of course, every memory function represents a different evolution rule but with elements of the original rule.

\subsection{Computing Potential}

In \cite{kn:MAC12} we show how a number of solitonic collisions can be simulated in rule 54. These solitons can be manipulated to develop some basic computable systems, such as simple substitution systems. In \cite{kn:MAM06} basic logic functions were simulated from basic collisions in rule 54. So far no one has ever implemented an equivalent Turing machine in rule 54. However, taking advantage of codification of gliders in rule 54, we have explored some basic computable functions that could help us to emulate the Turing machine with rule 54 in the future.

Some series by reaction gliders are presented in \cite{kn:Wolf02}. Here we have three cases.

\begin{figure}
\centerline{\includegraphics[width=2in]{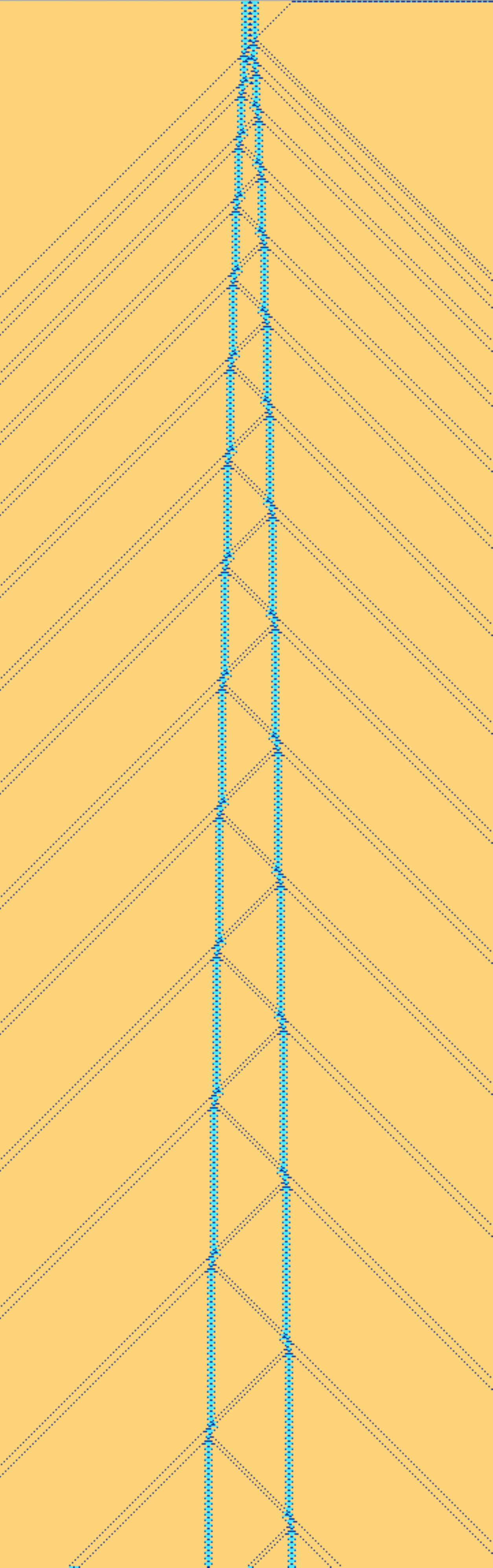}}
\caption{ECA rule 54 evolution deriving a series that yield $\mathbb{Z}^n$ for $n>2$.}
\label{series1}
\end{figure}

\begin{figure}
\centerline{\includegraphics[width=2.2in]{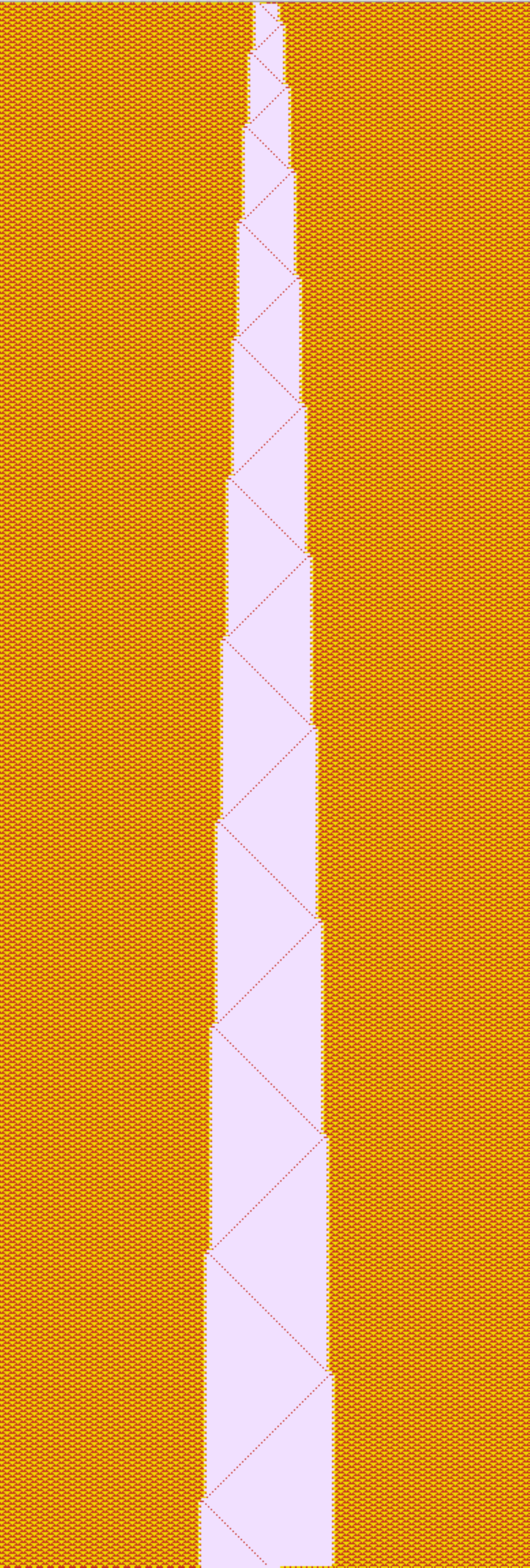}}
\caption{ECA rule 54 evolution deriving a parity function.}
\label{series2}
\end{figure}

\begin{figure}
\centerline{\includegraphics[width=1.8in]{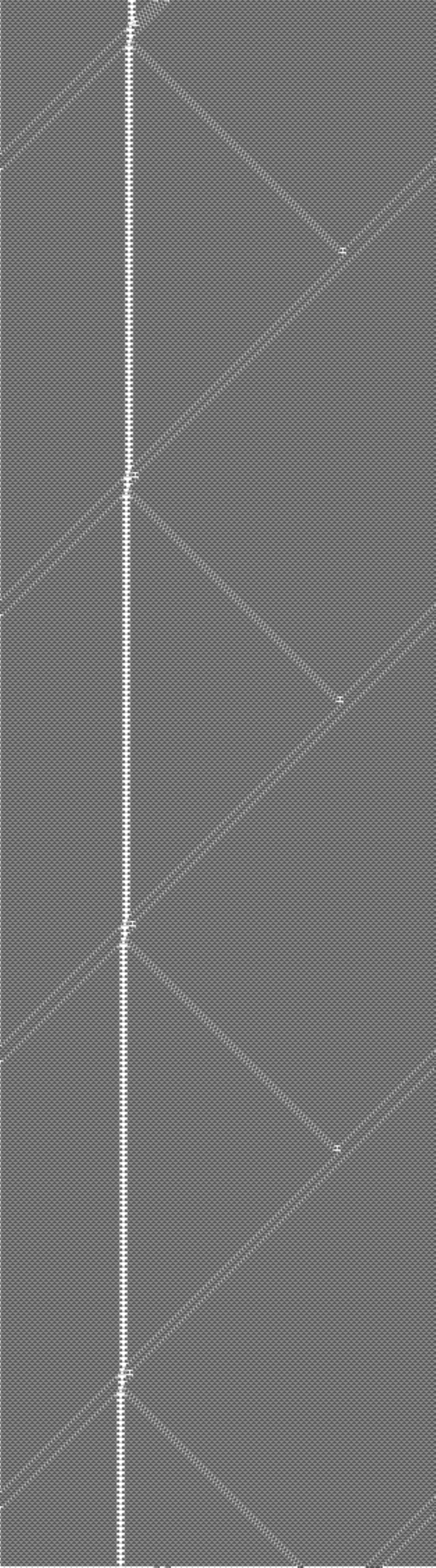}}
\caption{ECA rule 54 evolution implements a simple flip-flop.}
\label{flipflop}
\end{figure}

\begin{enumerate}
\item $\mathbb{Z}^n$ $\forall$ $n>3$ = $ne_{1}$-$2g_e$(A,f$_1$)-6$e_{1}$-$\overleftarrow{w}$-$ne_{2}$ (Figure~\ref{series1}).
\item Parity = $n$[$g_o$(A,f$_1$)-$g_o$(B,f$_1$)]-2$e_{1}$-$\overrightarrow{w}$-$2e_{2}$-$n$[$g_o$(A,f$_2$)-$g_o$(B,f$_2$)] (Figure~\ref{series2}).
\item Flip-flop = $ne_{1}$-$g_e$(A,f$_1$)-2$e_{1}$-$4\overleftarrow{w}$-$ne_{1}$ (Figure~\ref{flipflop}).
\end{enumerate}

In Figure~\ref{series1}, starting from a collision among three gliders yields an infinite series $\mathbb{Z}^n$ for $n>2$ (without limit boundaries). This sequence is defined by vertical number of $T_6$ tiles without some perturbation that evolves on each collision. Figure~\ref{series2} displays an evolution that simulates a parity function $2k$ $\forall$ $k \in \mathbb{Z}$. This parity is preserved by number of generations or by number of $T_5$ tiles ($g_o$ gliders) (without limit boundaries). So, Figure~\ref{flipflop} shows a very simple flip-flop configurations that is restricted to limit boundaries. All previous simulations needs more than 1000 generations.

\section{Final Remarks}

Cellular automaton gliders are analogs of optical solitons, kinks in polymer chains, excitation in molecular arrays (reaction diffusion computers \cite{kn:ACA05}, wave packets used slime mould to communicate information to distant part of the body \cite{kn:Ada10}), and defects in micro-tubules~\cite{kn:Ada02}. Also, rule 54 \emph{per se} is a discrete analog an active nonlinear medium with lateral inhibition between micro-volumes. The lateral inhibition in the nervous system sharpens and strengthens sensor perception and is widely employed in vision and olfactory systems. Thus we can speculate that the rule 54 is a simplest abstract model of the affective nervous system. The gliders then play a role of propagating action potential wave packets, and glider guns symbolize activity in the sources of sensorial stimulation. As we can see, there are many analogies of rule 54 behavior in physical and biological systems. And therefore, behavior of these systems can be described by unique subsets of regular expressions, where phase, distance, momentum, position, period, and speed are taken into consideration.



\begin{thebibliography}{99}

\bibitem{kn:Wolf94} S. Wolfram, {\em Cellular Automata and Complexity: Collected Papers}, Reading, MA: Addison-Wesley, 1994.

\bibitem{kn:Mar13} G. J. Mart{\'i}nez, ``A Note on Elementary Cellular Automata Classification,'' {\em Journal of Cellular Automata}, {\bf 8}(3-4), 2013 pp. 233--259.

\bibitem{kn:BNR91} N. Boccara, J. Nasser, and M. Roger, ``Particlelike Structures and their Interactions in Spatiotemporal Patterns Generated by One-Dimensional Deterministic Cellular Automaton Rules,'' {\em Physical Review A}, {\bf 44}(2), 1991 pp. 866--875. doi:10.1103/PhysRevA.44.866.

\bibitem{kn:HC97} J. E. Hanson and J. P. Crutchfield, ``Computacional Mechanics of Cellular Automata: An Example,'' {\em Physics D: Nonlinear Phenomena}, {\bf 103}(1-4), 1997 pp. 169--189. dii:10.1016/S0167-2789(96)00259-X.

\bibitem{kn:Wue11} A. Wuensche, (2011) {\em Exploring Discrete Dynamics}, Frome, England: Luniver Press, 2011.

\bibitem{kn:Wolf02} S. Wolfram, {\em A New Kind of Science}, IL: Wolfram Media, Inc., 2002.

\bibitem{kn:Mar00} B. Martin, ``A Group Interpretation of Particles Generated by One-Dimensional Cellular Automaton, Wolfram's Rule 54,'' {\em International Journal of Modern Physics C}, {\bf 11}(1), 2000 pp. 101--123.

\bibitem{kn:MAM06} G. J. Mart{\'i}nez, A. Adamatzky, and H. V. McIntosh,  ``Phenomenology of Glider Collisions in Cellular Automaton Rule 54 and Associated Logical Gates,'' {\em Chaos, Solitons \& Fractals}, {\bf 28}(1), 2006 pp. 100--111. doi:10.1016/j.chaos.2005.05.013.

\bibitem{kn:Guan12} J. Guan, ``Complex Dynamics of the Elementary Cellular Automaton Rule 54,'' {\em International Journal of Modern Physcis C}, {\bf 23}(7), 2012 p. 1250052. doi:10.1142/S0129183112500520.

\bibitem{kn:Red10} M. Redeker, ``Gliders and Ether in Rule 54,'' arxiv.org/abs/1007.2920v1.

\bibitem{kn:MAC12} G. J. Mart{\'i}nez, A. Adamatzky, F. Chen, and L. Chua, ``On Soliton Collisions between Localizations in Complex Elementary Cellular Automata: Rules 54 and 110 and Beyond,'' {\em Complex Systems}, {\bf 21}(2), 2012 pp. 117--142. \url{http://www.complex-systems.com/pdf/21-2-2.pdf}.

\bibitem{kn:MAA13} G. J. Mart{\'i}nez, A. Adamatzky, and R. Alonso-Sanz,  ``Designing Complex Dynamics in Cellular Automata with Memory,'' {\em International Journal of Bifurcation and Chaos}, {\bf 23}(10), 2013 p. 1330035. doi:10.1142/S0218127413300358.

\bibitem{kn:MAM08} G. J. Mart{\'i}nez, A. Adamatzky, and H. V. McIntosh,  ``On the Representation of Gliders in Rule 54 by de Bruijn and Cycle Diagrams,'' {\em Lecture Notes in Computer Science}, {\bf 5191} 2008 pp. 83--91. doi:10.1007/978-3-540-79992-4\_11.

\bibitem{kn:repR54} G. J. Mart{\'i}nez. ``Elementary Cellular Automaton Rule 54.'' (Jul. 11, 2014)  \url{http://uncomp.uwe.ac.uk/genaro/Rule54.html}.

\bibitem{kn:GS82} B. Gr\"{u}nbaum and G. C. Shephard, {\em Tilings and Patterns}, New York: W. H. Freeman and Company, 1987.

\bibitem{kn:Mc91} H. V. McIntosh, ``Linear Cellular Automata via de Bruijn Diagrams,'' (Jul 11, 2014) \url{http://delta.cs.cinvestav.mx/~mcintosh/oldweb/pautomata.html}.

\bibitem{kn:Mc09} H. V. McIntosh, {\em One Dimensional Cellular Automata}, Beckington, UK: Luniver Press, 2009.

\bibitem{kn:WL92} A. Wuensche and M. Lesser, {\em The Global Dynamics of Cellular Automata}, Reading: Addison-Wesley Publishing Company, 1992.

\bibitem{kn:Mc99} H. V. McIntosh, ``Rule 110 as it Relates to the Presence of Gliders,'' (Jul 23, 2014) \url{http://delta.cs.cinvestav.mx/~mcintosh/comun/RULE110W/RULE110.html}.

\bibitem{kn:MMS06} G. J. Mart{\'i}nez, H. V. McIntosh, and J. C. Seck-Tuoh-Mora, ``Gliders in Rule 110,'' {\em International Journal of Unconventional Computing}, {\bf 2}(1), 2006 pp. 1--49. \url{http://uncomp.uwe.ac.uk/genaro/Papers/Papers_on_CA_files/MARTINEZ.pdf}.

\bibitem{kn:Mc00} H. V. McIntosh, ``A Concordance for Rule 110,'' (Jul 11, 2014). \url{http://delta.cs.cinvestav.mx/~mcintosh/oldweb/pautomata.html}.

\bibitem{kn:MMS08} G. J. Mart{\'i}nez, H. V. McIntosh, J. C. Seck-Touh-Mora, and S. V. Chapa-Vergara, ``Determining a Regular Language by Glider-Based Structures Called Phases f$_i$\_1 in Rule 110,'' {\em Journal of Cellular Automata}, {\bf 3}(3), 2008 pp. 231--270.

\bibitem{kn:Voor96} B. H. Voorhees, {\em Computational Analysis of One-Dimensional Cellular Automata}, River Edge, NJ: World Scientific, 1996.

\bibitem{kn:McNX} H. V. McIntosh. ``Cellular Automata Packages.'' (Jul 11, 2014) \url{http://delta.cs.cinvestav.mx/~mcintosh/oldweb/software.html}.

\bibitem{kn:MarOSX} G. J. Mart{\'i}nez. ``OSXCA Systems.'' (Jul 11, 2014) \url{http://uncomp.uwe.ac.uk/genaro/OSXCASystems.html}.

\bibitem{kn:Voor08} B. H. Voorhees, ``Remarks on Applications of  De Bruijn Diagrams and Their Fragments,'' {\em Journal of Cellular Automata} {\bf 3}(3), 2008 187--204.

\bibitem{kn:HU79} J. E. Hopcroft and J. D. Ullman, {\em Introduction to Automata Theory Languages, and Computation}, Reading: Addison-Wesley Publishing  Company, 1979.

\bibitem{kn:Mins67} M. Minsky, {\em Computation: Finite and Infinite Machines}, Englewood Cliffs, NJ:Prentice Hall, 1967.

\bibitem{kn:agar} Wikipedia. ``Agar.'' (Jul 11, 2014) \url{http://conwaylife.com/wiki/Agar}.

\bibitem{kn:chess} Wikipedia. ``Chess Programming: Bill Gosper.'' (Jul 30, 2014) \url{http://chessprogramming.wikispaces.com/Bill+Gosper}.

\bibitem{kn:Kau93} S. A. Kauffman, {\em The Origins of Order: Self-Organization and Selection in Evolution}, New York: Oxford University Press, 1993.

\bibitem{kn:ACA05} A. Adamatzky, B. L. Costello, and T. Asai, {\em Reaction-Diffusion Computers}, Boston: Elsevier, 2005.

\bibitem{kn:Ada10} A. Adamatzky, {\em  Physarum Machines: Computers from Slime Mould}, Hackensack, NJ: World Scientific Publishing Co. Pte. Ltd. 2010.

\bibitem{kn:Ada02} A. Adamatzky, ed. {\em Collision-Based Computing}, London: Springer, 2002.

\end{thebibliography}
\end{document}